\documentclass{aa}

\usepackage[varg]{txfonts}
\usepackage{natbib,twoopt}
\usepackage{rotating}
\usepackage{float}

\newcommand{\giraffe}{{FLAMES\slash GIRAFFE}}

\newcommand{\fei}{\hbox{Fe\,{\scriptsize I}}}
\newcommand{\feii}{\hbox{Fe\,{\scriptsize II}}}

\newcommand{\cai}{\hbox{Ca\,{\scriptsize I}}}

\newcommand{\oi}{\hbox{O\,{\scriptsize I}}}
\newcommand{\nai}{\hbox{Na\,{\scriptsize I}}}
\newcommand{\mgi}{\hbox{Mg\,{\scriptsize I}}}
\newcommand{\sii}{\hbox{Si\,{\scriptsize I}}}
\newcommand{\tii}{\hbox{Ti\,{\scriptsize I}}}
\newcommand{\tiii}{\hbox{Ti\,{\scriptsize II}}}

\newcommand{\moog}{{\scriptsize MOOG}}
\newcommand{\marcs}{{\scriptsize MARCS}}
\newcommand{\vald}{{\scriptsize VALD}}

\newcommand{\msun}{\hbox{M$_{\odot}$}}

\newcommand{\teff}{\hbox{T$_{\mathrm{eff}}$}}
\newcommand{\logg}{\hbox{$\log g$}}
\newcommand{\feh}{\hbox{[Fe/H]}}

\newcommand{\xh}{\hbox{[X/H]}}
\newcommand{\kms}{\hbox{km\,s$^{-1}$}}

\newcommand{\feih}{\hbox{[\fei/H]}}

\newcommand{\nah}{\hbox{[Na/H]}}
\newcommand{\mgh}{\hbox{[Mg/H]}}

\newcommand{\cah}{\hbox{[Ca/H]}}

\newcommand{\ofe}{\hbox{[O/Fe]}}
\newcommand{\nafe}{\hbox{[Na/Fe]}}
\newcommand{\mgfe}{\hbox{[Mg/Fe]}}
\newcommand{\sife}{\hbox{[Si/Fe]}}
\newcommand{\cafe}{\hbox{[Ca/Fe]}}

\newcommand{\oife}{\hbox{[O\,{\scriptsize I}/Fe]}}
\newcommand{\naife}{\hbox{[Na\,{\scriptsize I}/Fe]}}
\newcommand{\mgife}{\hbox{[Mg\,{\scriptsize I}/Fe]}}
\newcommand{\siife}{\hbox{[Si\,{\scriptsize I}/Fe]}}
\newcommand{\caife}{\hbox{[Ca\,{\scriptsize I}/Fe]}}
\newcommand{\tiiife}{\hbox{[Ti\,{\scriptsize II}/Fe]}}

\newcommand{\afe}{\hbox{[$\alpha$/Fe]}}

\newcommand{\uu}{\hbox{\it U\/}}
\newcommand{\bb}{\hbox{\it B\/}}
\newcommand{\vv}{\hbox{\it V\/}}

\newcommand{\ii}{\hbox{\it I\/}}

\newcommand{\umb}{\hbox{\uu--\bb\/}}

\newcommand{\bmv}{\hbox{\bb--\vv\/}}
\newcommand{\bmi}{\hbox{\bb--\ii\/}}

\newcommand{\cubi}{\hbox{{\it c}$_{\rm U,B,I}$\/}}

\begin{document}

\title{The Carina Project. VIII. The $\alpha$-element abundances.
\thanks{Based on spectra retrieved from the ESO/ST-ECF Science 
Archive Facility and collected either with UVES at ESO/VLT (065.N-0378(A),
066.B-0320(A), P.I.: E.~Tolstoy) or with 
\giraffe-UVES at ESO/VLT (074.B-0415(A), 
076.B-0146(A), P.I.: E.~Tolstoy; 171.B-0520(A)(B)(C), 
180.B-0806(B), P.I.: G.~Gilmore).}$^,$
\thanks{Tables 2-6 and 8 are entirely in electronic form
at the CDS via anonymous ftp to cdsarc.u-strasbg.fr (130.79.128.5)
or via http://cdsweb.u-strasbg.fr/cgi-bin/qcat?J/A+A/.}
}
 
\author{M.~Fabrizio\inst{1}\and
M.~Nonino\inst{2}\and
G.~Bono\inst{3,4}\and
F.~Primas\inst{5}\and
F.~Th\'evenin\inst{6}\and
P.~B.~Stetson\inst{7}\and
S.~Cassisi\inst{1}\and
R.~Buonanno\inst{1,3}\and
G.~Coppola\inst{8}\and
R.~O.~da~Silva\inst{3}\and
M.~Dall'Ora\inst{8}\and
I.~Ferraro\inst{4}\and
K.~Genovali\inst{3}\and
R.~Gilmozzi\inst{5}\and
G.~Iannicola\inst{4}\and
M.~Marconi\inst{8}\and
M.~Monelli\inst{9,10}\and
M.~Romaniello\inst{5}\and
A.~R.~Walker\inst{11}
}

\institute{
INAF-Osservatorio Astronomico di Teramo, Via Mentore Maggini s.n.c., I-64100 Teramo, Italy\\
\email{fabrizio@oa-teramo.inaf.it}
\and INAF-Osservatorio Astronomico di Trieste, Via G.B. Tiepolo 11, I-40131 Trieste, Italy
\and Dipartimento di Fisica, Universit\`a di Roma "Tor Vergata", Via della Ricerca Scientifica 1, I-00133 Roma, Italy
\and INAF-Osservatorio Astronomico di Roma, Via Frascati 33, I-00040 Monte Porzio Catone (RM), Italy
\and European Southern Observatory, Karl-Schwarzschild-Str. 2, G-85748 Garching bei Munchen, Germany
\and Universit\'{e} de Nice Sophia-antipolis, CNRS, Observatoire de la C\^{o}te d'Azur, Laboratoire Lagrange, BP 4229, F-06304 Nice, France
\and Dominion Astrophysical Observatory, NRC-Herzberg, National Research Council, 5071 West Saanich Road, Victoria, BC V9E 2E7, Canada
\and INAF-Osservatorio Astronomico di Capodimonte, Salita Moiariello 16, I-80131 Napoli, Italy
\and Instituto de Astrof\'{i}sica de Canarias, Calle Via Lactea s/n, E-38200 La Laguna, Tenerife, Spain
\and Departamento de Astrof\'{i}sica, Universidad de La Laguna, E-38200 La Laguna, Tenerife, Spain
\and Cerro Tololo Inter-American Observatory, National Optical Astronomy Observatory, Casilla 603, La Serena, Chile
}

\date{Received .../ Accepted ...}

\abstract{We have performed a new abundance analysis of Carina red giant (RG) stars
from spectroscopic data collected with UVES (high spectral resolution)
and \giraffe\ (high and medium resolution) at ESO/VLT. The former sample
includes 44 RGs, while the latter consists of 65 (high-resolution) and $\sim$800
(medium-resolution) RGs, covering a significant fraction of the galaxy's
RG branch, and red clump stars. To improve the abundance analysis at the faint
magnitude limit, the \giraffe\ data were divided into ten
surface gravity and effective temperature bins. 
The spectra of the stars belonging to the same gravity and temperature bin 
were stacked. This approach allowed us to increase the signal-to-noise ratio 
in the faint magnitude limit (\vv$\ge$20.5~mag)
by at least a factor of five.
We took advantage of the new photometry index \cubi\ introduced recently 
as an age and probably a metallicity indicator to split stars along the red giant branch.
These two stellar populations display 
distinct \feh\ and \mgh\ distributions: their mean iron abundances
are --2.15$\pm$0.06~dex ($\sigma$=0.28), and --1.75$\pm$0.03~dex 
($\sigma$=0.21), respectively.
The two iron distributions differ at the 75\%\ level. This supports
preliminary results. 
Moreover, we found that the old and intermediate-age stellar populations
have mean \mgh\ abundances of --1.91$\pm$0.05~dex ($\sigma$=0.22)
and --1.35$\pm$0.03~dex ($\sigma$=0.22); these differ at the 83\%\ level.
Carina's $\alpha$-element abundances agree, within 1$\sigma$,
with similar abundances for field halo stars and for cluster (Galactic and
Magellanic) stars. The same outcome applies to nearby dwarf spheroidals
and ultra-faint dwarf galaxies in the iron range covered by Carina
stars.
Finally, we found evidence of a clear correlation between
Na and O abundances, thus suggesting that Carina's chemical enrichment
history is quite different from that in the globular clusters.
}


\authorrunning {Fabrizio et al.}
\titlerunning {Carina Project. VIII. The $\alpha$-element abundances.}

\maketitle

\section{Introduction}
\label{sec:intro}

\begin{figure*}
\centering
\includegraphics[trim=0cm 1.5cm 1cm 15.5cm,clip,width=1.0\textwidth]{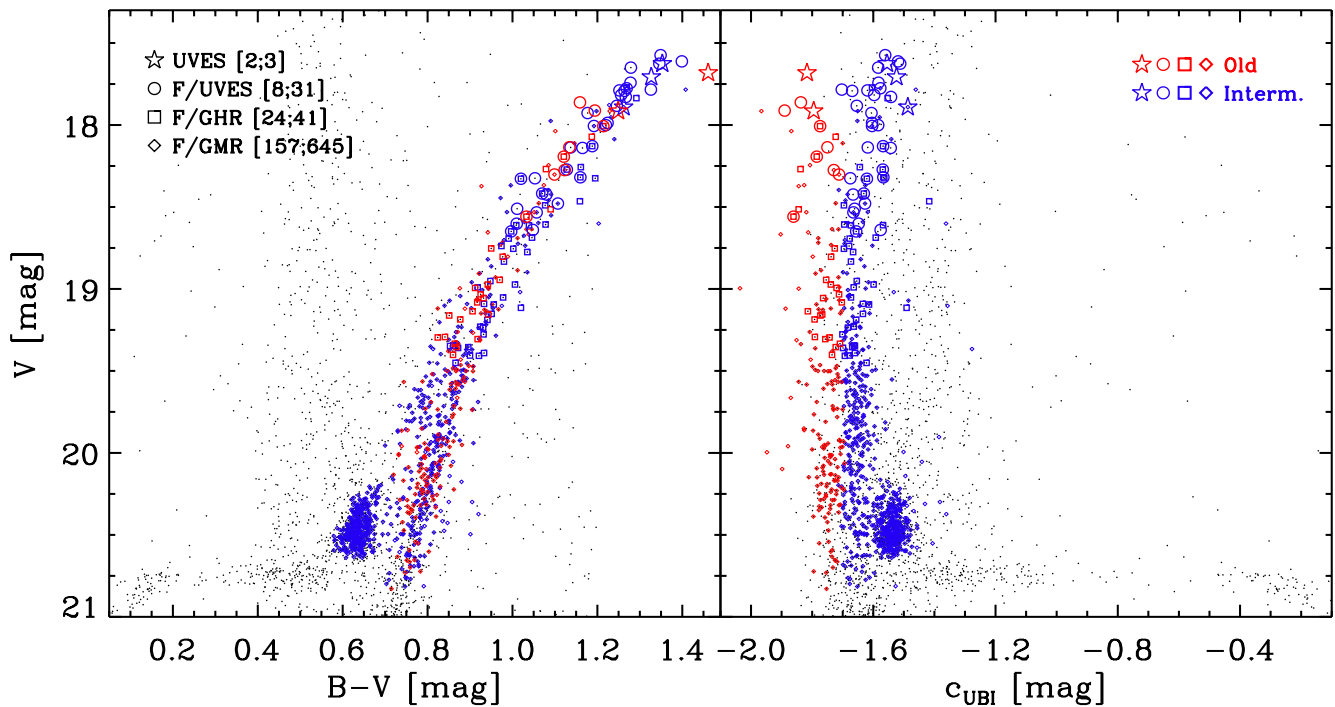}
\caption{Left: the brighter portion of the CMD of the Carina dwarf in the \vv\ vs \bmv\ plane. The
large colored circles represent the spectroscopic targets
in this investigation. The color coding derives from the selection
criteria shown in the right panel. Right: \vv\ vs \cubi\ diagram for the
same stars. This particular color combination,
(\umb)--(\bmi), allows us to split the RGB into old and intermediate-age
populations \citep{monelli14}. \label{cmdcubi}}
\end{figure*}

Empirical evidence indicates that dwarf spheroidal galaxies (dSphs) and
ultra-faint dwarfs (UFDs) are the smallest stellar systems to be dominated by
dark matter (DM). This finding is supported by new and more
precise kinematic measurements \citep{walker09a,walker09b}, 
implying that dSphs and UFDs can provide
firm constraints on the smallest DM halos that can retain baryons. The
nearby systems have the added advantage that we can sample a significant 
fraction of their stellar content. Therefore, these interesting stellar 
systems offer a unique opportunity to simultaneously probe their stellar 
content and their total mass budget \citep{walker09b}.

There is intriguing empirical evidence that 
both low- and high-mass galaxies follow the stellar
mass-metallicity relation \citep{tinsley79}. However, current
extragalactic surveys indicate that large galaxies have flat gas-phase
metallicity gradients \citep{moran12}. Dwarf galaxies, instead, show 
different peaks in the metallicity distribution 
(e.g., Tucana, \citealt{monelli10tucana} and 
Sculptor, \citealt{deboer12}), but still lack firm evidence of a 
metallicity gradient \citep{vanzee06}. The available
evidence seems to suggest not only that dwarf galaxies appear
to be less efficient star formers, but also that their chemical
enrichment might have been different from that of massive
galaxies. 

Cosmological models also suggest that dSphs and UFDs are the fossil
records of the Galactic halo \citep{helmi08}. 
Therefore, their kinematic and chemical
properties can provide firm constraints on the formation and evolution
of the Milky Way (MW). However, recent measurements from
high-resolution spectra indicate that the $\alpha$-element abundances in
nearby dSphs are, for iron abundances larger than \feh$>$--2, typically
less enhanced than halo stars and Galactic globular clusters
where \afe$\approx$0.4 \citep{tolstoy09}. This conclusion is supported by a
recent investigation based on medium-resolution spectra collected with
X-Shooter at VLT for seven either extremely (\feh$<$--3) or very
(\feh$<$--2) metal-poor stars: \citet{starkenburg13} found that the 
$\alpha$ enhancement is similar in the mean to halo stars of similar metallicities, 
but the spread around the mean is larger than around the halo stars.

Spectroscopic measurements of metal-poor stars in UFDs support
the same scenario, and indeed \citet{gilmore13}, using
high-resolution spectra for seven very metal-poor red giants (RGs) in
Bo\"{o}tes~I, found that their $\alpha$ enhancement is consistent with
halo stars, but showing a spread around the mean. 

These findings indicate that the chemical enrichment in low-mass
dwarfs has been slower than in the Galactic halo \citep{cayrel04} and in
the Galactic bulge \citep{lagioia14}. This means that dwarf galaxies
might have played a minor role in building up the Galactic spheroid 
\citep{leaman13,stetson14,fiorentino15}.

On the other hand, all the Galactic globular clusters (GCs) investigated
so far show a specific chemical fingerprint: the abundances of Na--O
and Mg--Al are anticorrelated \citep{carretta09uves,carretta09gir,carretta14}. 
This signature becomes even more compelling
if we consider the fact that field halo stars do not show evidence of
this anticorrelations \citep{gratton00}.
Moreover, massive GCs with a large spread in iron abundance but a clear 
evidence of Na--O and Mg--Al anticorrelations 
($\omega$-Cen, \citealt{johnson08}; 
M54, \citealt{carretta10m54}) have also been considered relic cores of
disrupted dwarf galaxies \citep{bekki03}. However, the current dwarf galaxies
for which we have a detailed knowledge of their chemical enrichment history
show a wide range in iron abundance, but no evidence of anticorrelations.
This evidence indicates that the role played by GCs and dwarf galaxies
in the early formation of the Galactic spheroid is still puzzling.

In this context, the Carina dSph can play a crucial role since it 
is relatively close (DM$_0$=20.10~mag, \citealt{coppola13}), 
it shows at least two clearly separated star-formation episodes, and 
a wide range in iron that covers at least 1.5~dex. 
The old stellar population has an age of 12~Gyr, while the 
intermediate-age has ages ranging from 4 to 8 Gyr \citep{monelli03}.  
In the investigation based on high-resolution (R$\sim$20,000) spectra
collected with FLAMES at VLT for 35 RGs, \citet{lemasle12} found that the
old stellar component in Carina is metal-poor (\feh$<$--1.5) and slightly
$\alpha$-enhanced (\mgfe$>$0). On the other hand, the intermediate-age
population is metal-intermediate (--1.5$<$\feh$<$--1.2) and shows a broad
spread in $\alpha$ enhancement. Indeed, the stars range from being
$\alpha$-poor (\mgfe$<$--0.3) to $\alpha$-enhanced (\mgfe$\sim$0.3).
These findings have been independently supported by the detailed star
formation history performed by \citet{deboer14}. They found
evidence of different age-metallicity relations and different trends in
the $\alpha$-element distributions between old- and intermediate-age
subpopulations.
More recently, \citet{vdberg15} used the star formation history
provided by \citet{deboer14} and found that specific sets of
cluster isochrones, covering a broad range in iron and in
$\alpha$-element abundances, take account of old horizontal branch (HB) 
stars and red clump (RC) stars in
Carina.
It is worth mentioning that these analyses are typically based on
stellar ages estimated by comparing the position of the stars in 
color-magnitude diagrams (CMDs)
with specific stellar isochrones. This approach is prone to
observational errors in distance determination, photometry, elemental
abundances, and interstellar reddening. It is also affected by
theoretical uncertainties as the efficiency of diffusive processes,
nuclear cross sections, and treatment of superadiabatic convection. For a more detailed discussion of the error budget we
refer to \citet{renzini91} and \citet{cassisi13book}.

A detailed spectroscopic analysis of Carina stars was also performed by 
\citet{venn12} using high-resolution (R$\sim$40,000) spectra for nine bright RGs, 
collected with UVES at VLT and with MIKE at Magellan. They found evidence of 
inhomogenous mixing between the old and the intermediate-age population. 
In particular, a broad spread in Mg was considered suggestive of poor mixing 
in the gas from which the old population formed, while the offset in 
$\alpha$-element abundance between the old and the intermediate-age population 
suggested that the second broader star formation episode in Carina took place in
$\alpha$-enriched gas.   

The present investigation of the chemical enrichment history of 
this interesting system is based on the largest homogeneous data 
set of Carina chemical abundances yet obtained.
Our motivation is twofold:\\
$(i)$ To distinguish old and intermediate-age Carina stars, we use the
\cubi=(\umb)--(\bmi) index \citep{monelli13,monelli14}. 
Detailed photometric
investigations indicate that this index can remove the degeneracy
between age and metallicity along the red giant branch (RGB). 
We note that one of the main
advantages of this index is that the separation of the two stellar
populations relies on a differential measurement. This means that it is
independent of uncertainties in the distance modulus, the
reddening, and the cluster isochrones.\\
$(ii)$ We secured high-resolution homogeneous spectra for 44 RGs
observed with either UVES or \giraffe-UVES with a nominal
spectral resolution of 40,000.
These spectra were supplemented with high- (R$\sim$20,000)
and medium-resolution (R$\sim$6,000) spectra collected with \giraffe.
Moreover, the latter spectra were also employed to investigate iron 
and $\alpha$ abundances down to the luminosity of the RC 
(\vv$\sim$20.5~mag).

The paper is organized as follows.
In Sect. \ref{sec:cubi} we introduce the photometric index \cubi\
and its use in separating the old and intermediate-age stellar populations
along the Carina RGB.
The three spectroscopic data sets adopted in the current investigation
are discussed in Sect. \ref{sec:observation}. In particular, we focus on spectral resolution, wavelength coverage, and the
signal-to-noise ratio of the different spectra.
In Sect. \ref{sec:stack} we describe the procedure adopted to
stack the \giraffe\ spectra in detail. This is a fundamental step for providing
accurate abundance determinations down to the RC magnitude level.
The techniques for measuring equivalent widths, for computing synthetic
spectra, and for estimating elemental abundances and their errors are
described in Sects. \ref{sec:ew} and \ref{sec:abund}. In these sections
we also present a comparison between the current results and those
available in the literature.
In Sect. \ref{sec:abund_vs_logg} we discuss the difference in iron and 
magnesium abundances between the old and intermediate-age Carina 
stellar populations.
The comparison between Carina's metallicity distribution and similar
abundances in Galactic halo stars and in Galactic and Magellanic globular
clusters are discussed in Sects. \ref{sec:abund_vs_mw} and
\ref{sec:abund_vs_GC}, respectively.
Comparisons between Carina's $\alpha$-element abundances and similar
abundances in dSph and UFD galaxies are presented 
in Sect. \ref{sec:abund_vs_dwarf}.
In Sect. \ref{sec:na-o} we investigate the possibile occurrence of
a correlation between Na and O abundances in Carina RGs.
Finally, in Sect.\ref{sec:summary} we summarize the results and outline the
future prospects of the Carina Project.

\section{ \cubi\ index and different stellar populations}
\label{sec:cubi}

\begin{figure}
\centering
\includegraphics[trim=2.3cm 1.5cm 8.cm 15.8cm,clip,width=0.95\columnwidth]{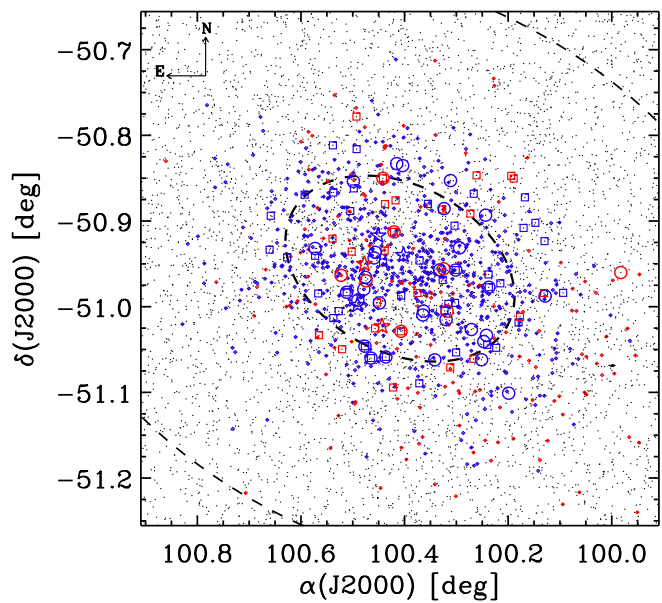}
\caption{Spatial distribution of our spectroscopic targets.
Symbols and colors are the same as in Fig.~\ref{cmdcubi}. The
dashed ellipses indicate the core and tidal radii of Carina
\citep{mateo98araa}. 
\label{mapcubi}} 

\end{figure}

\begin{figure*}[ht!]
\centering
\includegraphics[trim=0.cm 1.5cm 1.8cm 0.cm,clip,angle=90,width=0.9\textwidth]{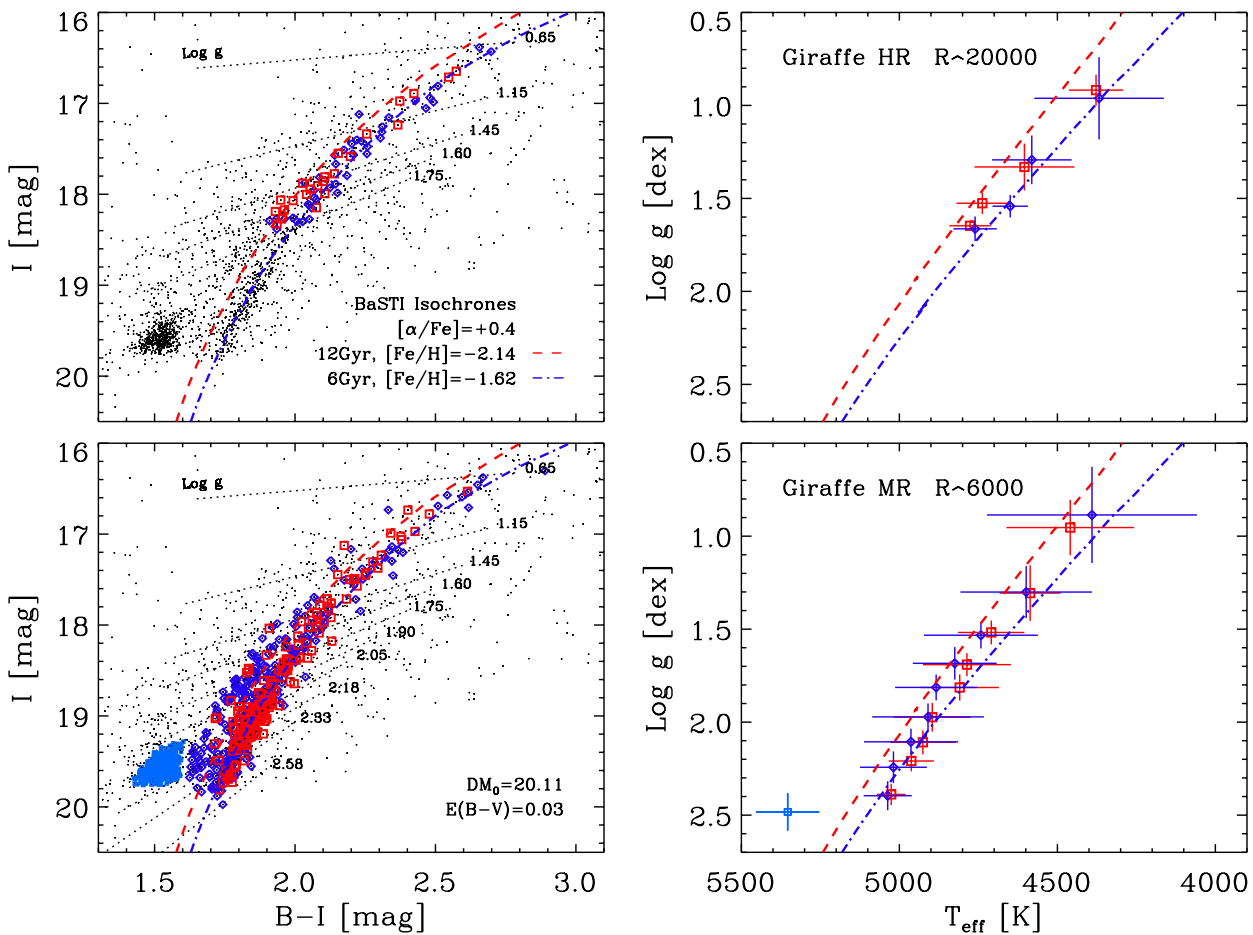}
\caption{Left column: \ii\ vs \bmi\ CMDs showing stars with
Giraffe HR (top) or MR (bottom) spectra. Red squares and blue diamonds show the
old and intermediate-age stellar components. Dotted lines mark the boundaries
of the gravity bins adopted for the spectrum-stacking procedure.
Two isochrones
from the BaSTI database \citep{pietrinferni04,pietrinferni06}
are shown as red dashed and blue dot-dashed lines. The adopted
distance modulus and reddening are indicated \citep{coppola13}. 
Right column: colored symbols show the positions of the stacked spectra
in the stellar parameter \logg\ vs \teff\ plane. \label{cmdstack}}
\end{figure*}

Recent results have revealed that different stellar populations in
old Galactic globular clusters can be easily isolated along the
whole CMD, from the main sequence, up to the subgiant branch,
RGB, and even the HB from an
appropriate combination of broadband filters
\citep{marino08,sbordone11,milone12}.
\citet{monelli13}
showed that their \cubi\ index is a powerful tool for identifying
multiple stellar sequences in the RGB of old GCs, and that the
\cubi\ pseudo-color of RGB stars correlates with the chemical
abundances of light elements. Moreover, \citet{monelli14} have shown
that \cubi\ can also distinguish a significant
fraction of the RGB stars of Carina's two main populations: the
old stars ($\sim$12~Gyr) have a more negative \cubi\ pseudo-color than
the intermediate-age stars (4-8~Gyr). 

Figure~\ref{cmdcubi} shows the \vv\ vs \bmv\ (left) and \vv\ vs
\cubi\ (right) diagrams for stars brighter than \vv=21~mag: the
brighter portion of Carina's RGB, the RC,
and part of the HB, and contaminating field stars at
\bmv$>$0.45~mag. We note that the main evolutionary features
in the \vv\ vs \cubi\ diagram are reversed, and the hottest stars attain
higher \cubi\ values. The distribution of Carina RGB stars in this
plane has been discussed by \citet{monelli14}, who showed that the
\cubi\ index largely removes the age-metallicity degeneracy
affecting the RGB stars. Following this analysis, the
right panel of Fig.~\ref{cmdcubi} shows a selection of old,
more metal-poor (red symbols) and intermediate-age, less
metal-poor stars (blue symbols). In particular, the red and blue
symbols identify stars with \cubi$<$--1.7~mag and
\cubi$>$--1.7~mag, respectively. We note that in the
classical \vv\ vs \bmv\ plane these stars are mixed along the RGB. 
The different symbols mark the position of the different spectroscopic 
data sets (see labels and the discussion in 
Sect.~\ref{sec:observation}).

The anonymous referee suggested that we discuss in more detail whether the
\cubi\ index is either an age or a metallicity indicator. The
empirical evidence suggests that the \cubi\ index is mainly an age
diagnostics, as has previously been discussed in detail by \citet{monelli14}.
However, we address this question below to further support 
the empirical framework we are developing concerning Carina stellar 
populations.

Dating back to the seminal investigation by \citet{smecker96},
it became clear that Carina experienced two clearly separated star formation 
episodes. However, optical and optical to near-infrared CMDs indicate that the two 
subpopulations overlap along the RGB. The \cubi\ pseudo-color distribution 
shows clear evidence of an asymmetric and possibly dichotomous distribution
of RGB stars. 
It is plausible to assume that this distribution is correlated with the 
difference in age of the two subpopulations. This is the reason why we 
associated the red and the blue RGB stars with the old- and intermediate-age 
subpopulations. However, we cannot exclude that the \cubi\ index is also 
affected by heavy element abundances. This means that the \cubi\ distribution 
might also be affected by a difference in CNO and/or in $\alpha$-element 
abundances. The main conclusion of this investigation, that is, the presence of 
two subpopulations that experienced two different chemical enrichment 
histories, is not affected by the intrinsic parameters affecting the \cubi\
index.

In passing we note that the cut adopted to split old- and
intermediate-age subpopulations was fixed according to the \cubi\
distribution. It is arbitrary, but quantitative tests indicate that
plausible changes in the cut do not affect the conclusions concerning
the metallicity distributions of the two main subpopulations.

Finally, we note that the age-metallicity pairs found for individual
Carina stars by \citet{deboer14} and by \citet{lemasle12} cannot be
recovered in this analysis. The theoretical reasons that led us to
overtake the fit with individual cluster isochrones, and in turn
individual age estimates of RGB stars, have been discussed in
\citet{monelli14} and in Sect. \ref{sec:intro}.

        \section{Observations and data analyses}
        \label{sec:observation}

Our data were collected with two
spectrographs mounted at the UT2 (Kueyen) at the Very Large
Telescope (VLT) of the European Southern Observatory (ESO). The
Fibre Large Array Multi Element Spectrograph (FLAMES; \citealt{pasquini02})
multi-object spectrograph was used to collect high- and medium-resolution
spectra with both the Ultraviolet and Visual Echelle
Spectrograph (UVES; \citealt{dekker00}) and GIRAFFE fiber modes.
Moreover, we also included in our analysis spectra collected with
the slit-mode of UVES. 

        \subsection{UVES and FLAMES\slash UVES spectra}
        
We present an extension of the analysis for the high-resolution
(R$\sim$40,000) UVES and FLAMES\slash UVES red-arm spectra presented in
\citet[][Paper~V]{fabrizio12}, where we obtained the \fei\ and
\feii\ abundances of 44 red giant Carina stars (hereafter UVES).
The stars in Fig.~\ref{cmdcubi} represent the UVES
targets (five stars), while the circles are for stars
observed with FLAMES/UVES (39 stars). The numbers in parentheses
indicate the number of stars belonging to the old (2+8) and
intermediate-age (3+31) populations, respectively, based on the \cubi\ index.
Figure~\ref{mapcubi} shows the spatial distribution of our
spectroscopic targets with the same color coding and symbols. The
data reduction, radial velocity (RV) measurements, and 
estimation of the stellar parameters of these spectra follow
the approach described in Paper~V. In particular, the
spectroscopic targets used in this analysis, with photometric,
astrometric, and stellar parameters, are listed in Table~1 of
Paper~V. 

        \subsection{\giraffe\ spectra}  
To increase the spectroscopic data set and cover the whole
extent of the RGB up to the intermediate-age RC helium-burning
region (\vv$\sim$20.5 and \bmv$\sim$0.6~mag), we included in our
analysis spectra collected with \giraffe. In particular,
we adopted both the high-
(HR10, HR13, and HR14A)\footnote{HR10: 5339$<$$\lambda$(\AA)$<$5619, R=19,800\\ 
HR13: 6120$<$$\lambda$(\AA)$<$6405, R=22,500\\ 
HR14A: 6308$<$$\lambda$(\AA)$<$6701, R=17,740}
and the medium-resolution (LR08)\footnote{LR08: 8206$<$$\lambda$(\AA)$<$9400, R=6,500} 
spectra that were presented by \citet{koch06},
\citet{lemasle12}, and \citet[][Paper~IV]{fabrizio11}.
The stars with high-resolution spectra were selected using the
following criteria:
{\it (i)} their radial velocities are within 4$\sigma$ from the Carina velocity 
peak (180$<$RV$<$260~\kms) and the precision on the individual RVs is 
better than 10 \kms\ (71 stars);
{\it (ii)} they have been measured in at least three photometric bands (\uu, \bb, \ii);
{\it (iii)} they have \bmi\ colors that are typical of RGB stars at the same apparent
\ii-band magnitudes ($\Delta$(\bmi)$\le$0.25~mag).
We obtained a sample of 65 out of the 71 stars. Almost 50\%\ (35) 
of the selected stars have previously been analyzed by \citet{lemasle12}. 
The others are used here for the first time to estimate iron and 
$\alpha$-element abundances. We note that selected stars adopted 
in the stacked spectra have between two to eight individual spectra. 
We refer to the end of Sect. \ref{sec:stack} for a more 
detailed discussion concerning the number of stars per stacked spectrum.

Similar criteria were also adopted to select 802 stars from the \giraffe\ 
medium-resolution sample. In particular, we obtained 483 stars along the 
RGB out of a sample of 529 candidate Carina stars (91\%). In the RC region 
we included 319 stars out of 407 candidate Carina stars (78\%). We 
excluded anomalous Cepheids and bright RC stars. The selected stars, adopted 
in the stacked spectra, have between two to 35 individual spectra. 
The reduction of these spectra follows the approach described in Paper~IV.

\subsubsection{High resolution}
\label{sec:stack_HR}
The HR spectroscopic targets (hereafter GHR) are shown as colored
squares in Figs.~\ref{cmdcubi} and \ref{mapcubi}. The old
population includes 24 stars, while that with intermediate-age stars
includes 41 objects. The top left panel of
Fig.~\ref{cmdstack} shows these stars in the 
\ii\ vs \bmi\ CMD (red squares and blue diamonds). 
Here, we overplotted
two isochrones (from the BaSTI database\footnote{\tt
http://www.oa-teramo.inaf.it/basti}, 
\citealt{pietrinferni04,pietrinferni06}), representing the
two main star-formation episodes of Carina. The adopted true
distance modulus and reddening values are from
\citet{coppola13} and are labeled in the figure, and we
used extinction coefficients from \citet{mccall04}. The
isochrones were used to divide the sample into four bins, using
iso-gravity loci (dotted lines). This approach produced four
subsamples of spectra that we stacked
because they have quite similar stellar parameters. 
The stellar parameters of each individual star were determined
following the procedure described in Paper~V. In Table~\ref{tab_param_stack}
we list the
mean values of effective temperature and surface gravity for each bin
with their uncertainties and in Col. 4 the number of individual 
stars per stacked spectrum. We note that the uncertainties in the different
bins are the standard deviations of the individual stellar parameters
summed in quadrature.
The stacking procedure is described in Sect.~\ref{sec:stack}.
The top right panel of Fig.~\ref{cmdstack} shows the
position of stacked spectra in the \teff\ vs \logg\ plane
(see also Table~\ref{tab_param_stack}). The
bars indicate the range of stellar parameters covered by
individual spectra; they range from $\Delta\logg$$\sim$0.1~dex,
$\Delta\teff$$\sim$50~K to $\sim$0.25~dex, $\sim$200~K. The
signal-to-noise ratio (S/N) of the individual spectra
ranges from $\sim$10 to $\sim$50 for the brightest targets.
This data set was also adopted by \citet{lemasle12} to investigate
the chemical abundances of 35 Carina RG stars. 
It is worth mentioning that the ranges in \logg\ and \teff\ 
covered by individual spectra belonging to the same gravity and
temperature 
bin allow us to provide accurate abundance estimates. Indeed, the quoted 
variations in \teff\ and \logg\ (see Sect. \ref{subsec:ab_uncert}) cause 
an uncertainty on individual abundances of about 0.15~dex. 

\begin{table}
\footnotesize
\caption{Stellar parameters of stacked spectra.}
\label{tab_param_stack}
\centering
\begin{tabular}{cccc}
\hline\hline
 {ID}&
 {\teff\ (K)}&
 {$\log g$}&
 {N$^*$} \\
  
\hline
HRold1 & 4378$\pm$98 & 0.92$\pm$0.13  &  4 \\
HRold2 & 4604$\pm$112 & 1.33$\pm$0.15  &  7\\
HRold3 & 4738$\pm$80 & 1.53$\pm$0.11  &  9 \\
HRold4 & 4776$\pm$78 & 1.65$\pm$0.10  &  4 \\
\hline
HRint1 & 4368$\pm$130 & 0.96$\pm$0.19  &  9 \\
HRint2 & 4580$\pm$92 & 1.29$\pm$0.13  & 14 \\
HRint3 & 4649$\pm$67 & 1.54$\pm$0.11  &  7 \\
HRint4 & 4760$\pm$71 & 1.66$\pm$0.11  & 11 \\
\hline
LRold1 & 4459$\pm$129 & 0.95$\pm$0.15 &  8 \\
LRold2 & 4586$\pm$79 & 1.31$\pm$0.14 & 15 \\
LRold3 & 4709$\pm$75 & 1.52$\pm$0.11 & 11 \\
LRold4 & 4786$\pm$100 & 1.69$\pm$0.11 & 14 \\
LRold5 & 4809$\pm$74 & 1.81$\pm$0.11 & 20 \\
LRold6 & 4896$\pm$78 & 1.97$\pm$0.11 & 27 \\
LRold7 & 4926$\pm$64 & 2.11$\pm$0.11 & 32 \\
LRold8 & 4962$\pm$61 & 2.21$\pm$0.10 & 18 \\
LRold9 & 5026$\pm$56 & 2.39$\pm$0.11 & 12 \\
\hline
LRint1 & 4390$\pm$189 & 0.89$\pm$0.21 & 14 \\
LRint2 & 4598$\pm$134 & 1.30$\pm$0.14 & 19 \\
LRint3 & 4741$\pm$104 & 1.53$\pm$0.11 & 20 \\
LRint4 & 4824$\pm$93 & 1.68$\pm$0.11 & 33 \\
LRint5 & 4884$\pm$90 & 1.81$\pm$0.11 & 61 \\
LRint6 & 4909$\pm$84 & 1.97$\pm$0.11 & 52 \\
LRint7 & 4963$\pm$91 & 2.11$\pm$0.11 & 50 \\
LRint8 & 5019$\pm$73 & 2.24$\pm$0.11 & 52 \\
LRint9 & 5037$\pm$63 & 2.40$\pm$0.11 & 25 \\
LRrc   & 5354$\pm$68 & 2.48$\pm$0.10 & 319 \\
\hline 
\end{tabular}

\end{table}

\subsubsection{Medium resolution}
\label{sec:sec:stack_LR}
We repeated the approach described above with the LR08
spectra (hereafter GMR). This data set is the
combination of two observing runs in 2003 (GMR03) and
2008 (GMR08). The details of these samples and their
combination were discussed in Paper~IV. We obtained 157 stars in the old and 645 stars
in the intermediate-age population.
The bottom left panel of
Fig.~\ref{cmdstack} shows the CMD and iso-gravity loci. 
The sample was
split into nine bins, plus a particular region enclosing the RC
stars. 
The stellar parameters and their uncertainties were estimated
following the same approach discussed in Sect.~\ref{sec:stack_HR} 
and listed in Table~\ref{tab_param_stack}.
The S/N of the individual spectra ranges from $\approx$10
to 50 for GMR03 (17$\lesssim$\vv$\lesssim$20.5~mag) and from
$\approx$5 to 15 for GMR08 (18.5$\lesssim$\vv$\lesssim$20.75~mag). 
The positions of the stacked spectra in the \teff\ vs \logg\ plane
are shown in the bottom right panel of
Fig.~\ref{cmdstack} (see also Table~\ref{tab_param_stack}),
where we obtained values of variations from
$\Delta\logg$$\sim$0.1~dex, $\Delta\teff$$\sim$50~K to $\sim$0.25,
$\sim$300~K. In this context, it is worth mentioning that the
GMR08 sample was previously used to constrain the kinematic
properties of Carina stars (Paper~IV). However, this is the first
time they are used to constrain the elemental abundances of RG
stars down to the RC magnitude level. 

For clarity in tracing back the identification of adopted
spectra and stars, Table~\ref{tab_allstars} gives in the first three
columns the position ($\alpha$, $\delta$) and the current ID, based on
Paper~V. Columns 4 and 5 give the IDs of the UVES and GHR samples,
while Cols. 6 and 7 report the IDs of the GMR03 and GMR08 samples. The star
IDs adopted by \citet{venn12} are listed in Col. 8. Moreover, in
Table~\ref{tab_allstars} we also list the same information for the
individual GHR and GMR spectra adopted in the stacking of different
effective temperature and surface gravity bins (see next section).

\begin{table*}
\footnotesize
\caption{List of cross-identified spectroscopic targets.}
\label{tab_allstars}
\centering
\begin{tabular}{cccccccc}
\hline\hline
{$\alpha$ (J2000)} &
{$\delta$ (J2000)} &
{Fabrizio+12} &
{UVES} &
{GHR} &
{GMR03} &
{GMR08} &
{Venn+12} \\
(deg) & (deg) \\
\hline
 099.9825 & --50.9602 &Car13& LG04c\_000951\tablefootmark{b}& \dots & \dots & \dots &\dots\\
 100.1285 & --50.9876 &Car14& LG04b\_004260\tablefootmark{b}& \dots & LG04b\_004260& \dots &\dots\\
 100.1990 & --51.1010 &Car15& LG04c\_000626\tablefootmark{b}& \dots & LG04c\_000626& \dots &\dots\\
 100.2374 & --50.9773 &Car16& CC\_09869& MKV0825\tablefootmark{c}& \dots & \dots &\dots\\
 100.2379 & --50.9624 &Car17& CC\_09400& MKV0780\tablefootmark{c}& LG04a\_003830& \dots &\dots\\
 100.2419 & --51.0334 &Car18& CC\_11388& \dots & LG04c\_006573& \dots &\dots\\
 100.2436 & --50.8932 &Car19& LG04a\_001826\tablefootmark{b}& \dots & \dots & \dots &Car--612\\
 100.2471 & --51.0408 &Car20& CC\_11560& \dots & \dots & \dots &\dots\\
 100.2513 & --51.0620 &Car21& CC\_12038& MKV1012\tablefootmark{c}& LG04c\_006479& \dots &\dots\\
 100.2709 & --51.0267 &Car22& LG04c\_000777\tablefootmark{b}& \dots & \dots & \dots &\dots\\
 100.2940 & --50.9314 &Car23& CC\_08447& \dots & \dots & \dots &\dots\\
 100.3013 & --50.9573 &Car24& CC\_09226& MKV0770\tablefootmark{c}& LG04a\_003844& \dots &\dots\\
 100.3113 & --50.8528 &Car25& UKV0524& \dots & LG04a\_002065& \dots &Car--524\\
 100.3145 & --51.0211 &Car26& CC\_11083& \dots & LG04c\_006621& \dots &\dots\\
\hline
\multicolumn{7}{c}{\bf HRold1}\\
 100.1770 & --51.0119 & \dots & \dots & MKV0914\tablefootmark{c}& LG04c\_004227& \dots &\dots\\
 100.3274 & --50.8866 & \dots & \dots & MKV0596\tablefootmark{c}& \dots & \dots &\dots\\
 100.4069 & --51.0288 &Car36& LG04d\_006628\tablefootmark{b}& MKV0948\tablefootmark{c}& \dots & \dots &\dots\\
 100.4417 & --50.8502 &Car40& CC\_06486& MKV0514\tablefootmark{c}& \dots & \dots &\dots\\
\hline
\end{tabular}

\tablefoot{This table is available entirety in a machine-readable form in the 
online journal.\\
\tablefoottext{a}{Star ID according to \citet{shetrone03}.}\\
\tablefoottext{b}{Star ID according to \citet{koch08}.}\\
\tablefoottext{c}{Star ID according to \citet{lemasle12}.}}
\end{table*}

        \section{Stacking procedure for \giraffe\ spectra}
        \label{sec:stack}

\begin{figure}[ht!]
\centering
\includegraphics[trim=0.3cm 1.1cm 1cm 1.5cm,clip,width=1\columnwidth]{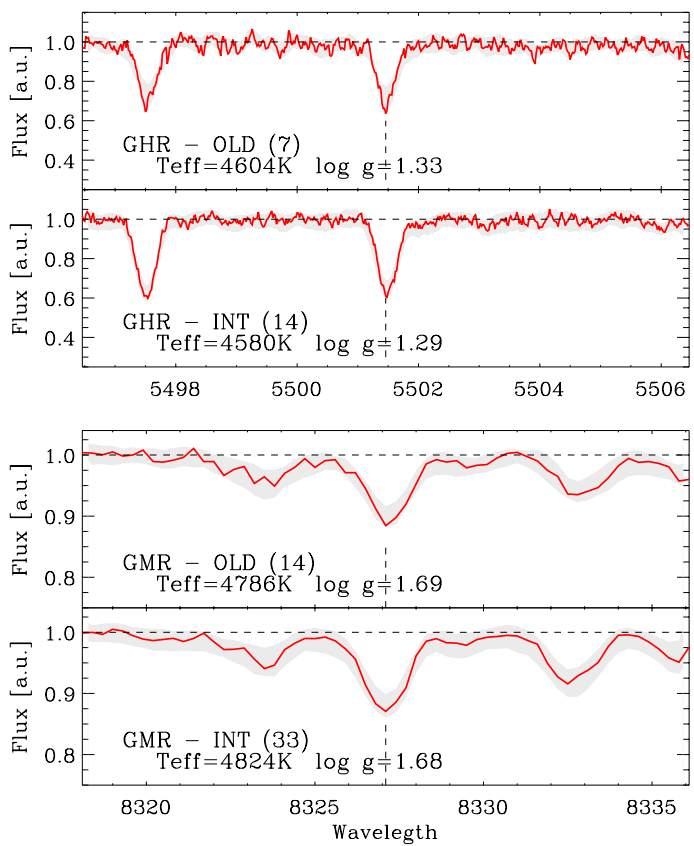}
\caption{Examples of stacked spectra. Top: resulting stack of 7(14)
spectra belonging to the old (intermediate) stellar population,
collected with the Giraffe HR10 grism. The shaded area shows the
dispersion of individual spectra. We show a portion around an
\fei\ line marked by the dashed line. Bottom: the same as the top
panel, but for spectra collected with Giraffe LR08 grism. 
\label{specstack}} 
\end{figure}

The individual spectra belonging to each bin and population were
stacked in a two-step procedure.\\ The first fundamental step is
estimating the continuum to gain individual normalized spectra.
By default, each spectrum is divided into 200 intervals. To
properly identify the continuum while avoiding lines, spikes, and
contaminants we calculated the biweight mean \citep{beers90} for
each interval using the inverse square-root of the signal as the
weight. The mean value was augmented by 75\%\ of the dispersion to
define the upper envelope of the signal. Then, the 200 local
estimates were connected using a running average with a fixed step
of 40. The resulting curve is a good approximation of the
continuum over the entire spectral range. The resulting normalized spectra can be visually checked
and, if the normalization
is problematic, the number of intervals and the averaging step
can be changed.\\ In the second step we averaged all normalized
spectra belonging to the different gravity bins of the two
populations. To do this, each spectrum was accurately rectified for
its radial velocity and then was rebinned with a fixed wavelength
step (depending on the resolution). Finally, a biweight mean was
applied to each wavelength step, averaging all spectra together. 
Stacking 4-9 (OLD) and 9-14 (INT) individual targets increased
the S/N of the GHR spectra, in particular for the faintest targets
in the last bin, by a factor of 3-4. For the GMR data set,
stacking 8-32 (OLD) and 14-61 (INT) individual targets increases
the S/N by a factor of 4-8 (see Table~\ref{tab_param_stack}). 
Figure~\ref{specstack} shows an example of the stacked spectrum
for an old and an intermediate-age star in the HR10 (top) and LR08
(bottom) grisms. The shaded area represents the dispersion of the
individual spectra, and the plots are centered on two \fei\ lines
that are recognizable in the wavelength range. 

        \section{Equivalent width measurement}
\label{sec:ew}
        \subsection{Line list and atomic data}
We selected isolated and unblended iron, sodium, and $\alpha$-element 
(\oi, \mgi, \sii, \cai, and \tiii) atomic 
lines in the wavelength range of our spectra from different sources in the literature. 
In particular, we merged 
the line lists of \citet{shetrone03}, \citet{koch08},
\citet{fabrizio12}, \citet{lemasle12}, and \citet{venn12}. We updated the
atomic data for these lines from the
\vald\footnote{\tt
http://www.astro.uu.se/$\sim$vald/php/vald.php} data base
\citep{kupka00}. 
The final line lists adopted for each data set are shown in 
the first four columns of Tables~\ref{tab_UVES}, \ref{tab_HR}, 
\ref{tab_LRold}, and \ref{tab_LRint}. They list the line wavelength (Col.1), 
element species (Col.2), excitation potential (Col.3), and $\log gf$ (Col.4).

\begin{figure*}
\centering
\includegraphics[trim=2.5cm 1.5cm 1cm 8.5cm,clip,width=0.8\textwidth]{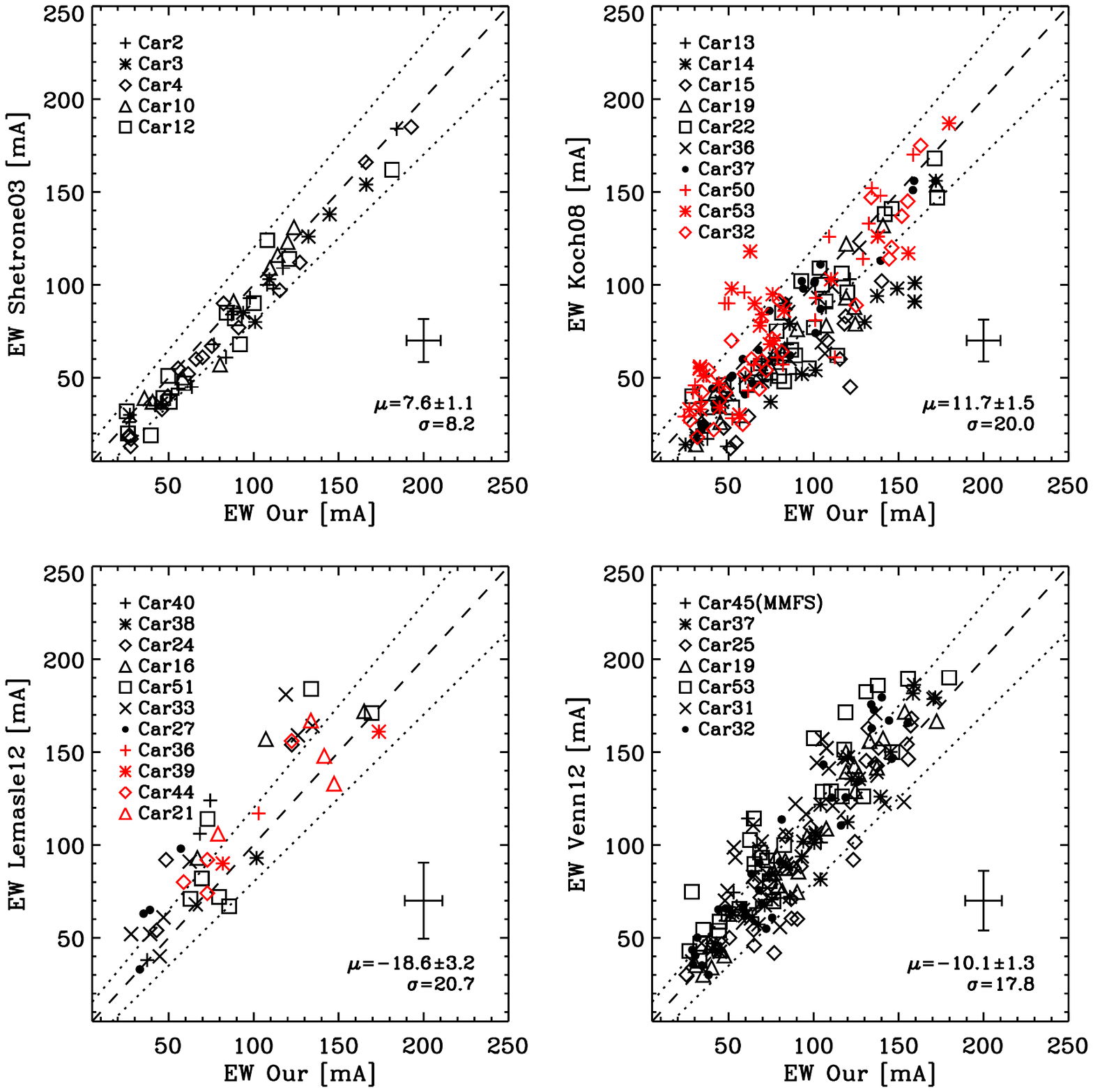}
\caption{Equivalent width comparisons for stars in common
with literature data sets: \citet{shetrone03} (top left), \citet{koch08}
(top right), \citet{lemasle12} (bottom left) and \citet{venn12} (bottom right). 
The dashed line shows the bisector of the plane. The dotted 
lines display the 10\%\ uncertainty convolved with a 10~m\AA\
error. The mean measurement errors are also displayed. \label{ew}}
\end{figure*}

\begin{table*}
\footnotesize
\caption{Equivalent widths (in m\AA) and their errors ($\epsilon$EW) for
individual UVES stars.}
\label{tab_UVES}
\centering
\begin{tabular}{cccc cccccc}
\hline\hline
 {$\lambda$ (\AA)}&
 {Elem.}&
 {$\chi$ (eV)}&
 {$\log gf$}&
   Car2 &    Car3 &    Car4 &   Car10 &   Car12 &   Car13 \dots \\
\hline
6300.304 &    \oi &   0.000 &  --9.819 &         55.9$\pm$6.6&       27.2$\pm$3.6&       49.5$\pm$7.2&             \dots &       46.8$\pm$5.8&            \dots \\
6363.776 &    \oi &   0.020 & --10.303 &               \dots &             \dots &       22.7$\pm$2.6&             \dots &       24.4$\pm$5.0&            \dots \\
5682.633 &   \nai &   2.102 &  --0.700 &        26.6$\pm$3.3&             \dots &             \dots &             \dots &             \dots &            \dots \\
5688.205 &   \nai &   2.104 &  --0.450 &        51.7$\pm$6.8&             \dots &       61.5$\pm$7.2&             \dots &       59.0$\pm$8.5&            \dots \\
6160.747 &   \nai &   2.104 &  --1.260 &         23.8$\pm$3.5&             \dots &             \dots &             \dots &       39.5$\pm$5.3&            \dots \\
5528.405 &   \mgi &   4.346 &  --0.620 &              \dots &     132.2$\pm$16.7&     182.6$\pm$19.4&     109.6$\pm$12.1&     182.1$\pm$19.3&      84.4$\pm$9.9\\
5711.088 &   \mgi &   4.346 &  --1.833 &       87.2$\pm$10.1&       45.4$\pm$5.5&      91.1$\pm$10.3&       40.5$\pm$5.0&       84.2$\pm$9.8&            \dots \\
5645.613 &   \sii &   4.930 &  --2.140 &              \dots &             \dots &       27.8$\pm$4.0&             \dots &             \dots &            \dots \\
5665.555 &   \sii &   4.920 &  --2.040 &              \dots &             \dots &       27.6$\pm$3.4&             \dots &       25.3$\pm$4.0&            \dots \\
5948.541 &   \sii &   5.082 &  --1.230 &              \dots &             \dots &       39.8$\pm$5.7&             \dots &             \dots &            \dots \\
6145.016 &   \sii &   5.616 &  --1.310 &              \dots &             \dots &       26.8$\pm$4.7&             \dots &             \dots &            \dots \\
6155.134 &   \sii &   5.619 &  --0.754 &              \dots &             \dots &       51.9$\pm$8.1&             \dots &       38.9$\pm$5.3&            \dots \\
6243.815 &   \sii &   5.616 &  --1.242 &              \dots &       24.7$\pm$4.2&             \dots &             \dots &             \dots &            \dots \\
6244.466 &   \sii &   5.616 &  --1.093 &              \dots &             \dots &             \dots &             \dots &       25.9$\pm$3.7&            \dots \\
5581.965 &   \cai &   2.523 &  --0.555 &       98.5$\pm$10.7&       82.6$\pm$9.3&     105.1$\pm$11.5&       42.7$\pm$6.1&      92.5$\pm$10.9&      37.5$\pm$6.7\\
5588.749 &   \cai &   2.526 &   0.358 &     155.2$\pm$16.7&             \dots &     152.1$\pm$16.0&      97.4$\pm$12.0&     149.0$\pm$16.1&     96.9$\pm$11.1\\
5590.114 &   \cai &   2.521 &  --0.571 &      110.3$\pm$14.3&       82.8$\pm$9.3&       94.1$\pm$9.9&       43.0$\pm$5.9&      86.9$\pm$10.8&      45.0$\pm$6.3\\
5601.277 &   \cai &   2.526 &  --0.523 &      108.2$\pm$13.0&      76.3$\pm$10.8&      94.3$\pm$10.4&             \dots &             \dots &      41.6$\pm$5.0\\
6122.217 &   \cai &   1.886 &  --0.316 &      184.1$\pm$20.4&     166.3$\pm$18.1&     192.7$\pm$20.5&     120.0$\pm$12.6&     181.4$\pm$19.1&    113.4$\pm$12.6\\
6161.297 &   \cai &   2.523 &  --1.266 &              \dots &             \dots &       66.1$\pm$7.3&             \dots &      91.9$\pm$12.3&            \dots \\
6162.173 &   \cai &   1.899 &  --0.090 &      202.7$\pm$21.8&     190.0$\pm$21.0&     199.0$\pm$20.7&     127.9$\pm$15.2&     196.8$\pm$20.7&    121.3$\pm$13.0\\
6166.439 &   \cai &   2.521 &  --1.142 &        76.6$\pm$9.3&             \dots &       69.9$\pm$7.5&             \dots &             \dots &      17.2$\pm$2.6\\
6169.042 &   \cai &   2.523 &  --0.797 &      106.1$\pm$13.0&       63.9$\pm$7.9&      95.7$\pm$10.6&             \dots &     107.2$\pm$13.5&      28.9$\pm$3.4\\
6169.563 &   \cai &   2.526 &  --0.478 &      115.8$\pm$13.6&      71.2$\pm$10.0&     101.6$\pm$10.9&             \dots &     113.0$\pm$12.4&      41.4$\pm$4.7\\
6439.075 &   \cai &   2.526 &   0.390 &             \dots &     144.7$\pm$16.9&     166.0$\pm$17.3&     114.1$\pm$13.4&             \dots &    107.3$\pm$11.7\\
6455.598 &   \cai &   2.523 &  --1.340 &        51.2$\pm$7.3&             \dots &       56.6$\pm$6.2&             \dots &       54.1$\pm$6.7&            \dots \\
6455.598 &   \cai &   2.523 &  --1.340 &        66.2$\pm$7.6&             \dots &       52.9$\pm$6.1&             \dots &       47.9$\pm$5.6&            \dots \\
6471.662 &   \cai &   2.526 &  --0.686 &      101.8$\pm$11.1&       72.7$\pm$9.8&             \dots &       31.6$\pm$5.8&       95.1$\pm$9.8&            \dots \\
6493.781 &   \cai &   2.521 &  --0.109 &              \dots &             \dots &             \dots &       68.8$\pm$7.5&             \dots &            \dots \\
6499.650 &   \cai &   2.523 &  --0.818 &       97.8$\pm$10.3&             \dots &       82.3$\pm$8.6&       35.7$\pm$4.9&             \dots &            \dots \\
6717.681 &   \cai &   2.709 &  --0.524 &      112.8$\pm$13.2&       65.4$\pm$9.8&      99.6$\pm$11.0&       43.3$\pm$5.3&     107.2$\pm$11.5&      38.4$\pm$4.6\\
4805.085 &  \tiii &   2.061 &  --0.960 &      109.6$\pm$11.6&     104.9$\pm$13.9&     112.0$\pm$12.8&      93.6$\pm$11.3&     121.2$\pm$13.9&            \dots \\
4911.193 &  \tiii &   3.124 &  --0.610 &        57.1$\pm$6.5&      61.7$\pm$11.0&             \dots &             \dots &             \dots &            \dots \\
5005.157 &  \tiii &   1.566 &  --2.720 &        55.0$\pm$5.7&       47.3$\pm$6.1&       57.8$\pm$8.4&             \dots &      74.4$\pm$10.4&            \dots \\
5154.068 &  \tiii &   1.566 &  --1.750 &      124.3$\pm$16.2&      97.6$\pm$13.2&     123.7$\pm$14.9&             \dots &     123.7$\pm$14.1&      55.9$\pm$9.8\\
5154.068 &  \tiii &   1.566 &  --1.750 &      107.4$\pm$11.3&     114.4$\pm$13.7&     132.2$\pm$16.3&      88.2$\pm$11.7&     119.5$\pm$13.2&      73.5$\pm$9.7\\
5185.902 &  \tiii &   1.893 &  --1.490 &      107.4$\pm$12.4&     117.6$\pm$15.3&             \dots &       64.8$\pm$9.3&     105.2$\pm$11.9&      54.7$\pm$6.4\\
5336.771 &  \tiii &   1.582 &  --1.582 &      138.2$\pm$17.6&     120.4$\pm$20.4&     135.7$\pm$17.2&     115.1$\pm$15.5&     135.6$\pm$19.2&     89.6$\pm$11.4\\
5336.771 &  \tiii &   1.582 &  --1.582 &      135.9$\pm$17.0&     136.8$\pm$17.6&             \dots &             \dots &     146.2$\pm$19.7&            \dots \\
5418.751 &  \tiii &   1.582 &  --2.000 &      111.8$\pm$13.9&      93.9$\pm$12.6&     115.3$\pm$12.7&      80.2$\pm$12.0&     100.0$\pm$13.1&      48.8$\pm$7.7\\

\hline
\end{tabular}

\tablefoot{This table is available entirety in a machine-readable form in the 
online journal.}
\end{table*}
\begin{table*}
\footnotesize
\caption{Equivalent widths (in m\AA) and their errors ($\epsilon$EW) for
stacked stars in Giraffe HR grisms.}
\label{tab_HR}
\centering
\begin{tabular}{cccc cccc cc}
\hline\hline
 {$\lambda$ (\AA)}&
 {Elem.}&
 {$\chi$ (eV)}&
 {$\log gf$}&
 {HRold1}&
 {HRold2}&
 {HRold3}&
 {HRold4}&
 {HRint1}&
 {HRint2 \dots}\\
\hline
\multicolumn{10}{c}{\bf Giraffe HR10}\\
5339.929& \fei&   3.266 &  --0.647 &\dots &81.7$\pm$13.4 &98.2$\pm$14.2 &99.5$\pm$17.4 &121.2$\pm$15.0 &128.1$\pm$16.7 \\
5341.024& \fei&   1.608 &  --1.953 &161.8$\pm$19.3 &122.8$\pm$16.7 &113.6$\pm$14.7 &95.7$\pm$15.3 &205.7$\pm$23.3 &156.9$\pm$18.5 \\
5367.466& \fei&   4.415 &   0.443 &63.0$\pm$10.5 &71.0$\pm$13.0 &73.4$\pm$11.3 &\dots &90.8$\pm$15.1 &74.4$\pm$12.8 \\
5383.369& \fei&   4.312 &   0.645 &75.0$\pm$10.2 &71.6$\pm$10.2 &57.9$\pm$9.0 &92.9$\pm$15.5 &113.7$\pm$13.3 &113.4$\pm$13.4 \\
5393.167& \fei&   3.241 &  --0.715 &88.9$\pm$11.8 &86.7$\pm$12.2 &74.5$\pm$11.0 &71.2$\pm$12.6 &124.4$\pm$16.8 &105.7$\pm$14.6 \\
5397.128& \fei&   0.915 &  --1.993 &178.8$\pm$21.2 &174.1$\pm$20.7 &168.3$\pm$21.0 &165.2$\pm$22.7 &226.0$\pm$26.2 &192.9$\pm$22.9 \\
5410.910& \fei&   4.473 &   0.398 &60.6$\pm$9.0 &52.8$\pm$8.9 &\dots &\dots &80.6$\pm$10.0 &77.1$\pm$10.2 \\
5429.696& \fei&   0.958 &  --1.879 &199.0$\pm$22.7 &179.4$\pm$21.3 &188.6$\pm$22.7 &156.6$\pm$20.8 &251.6$\pm$27.9 &226.6$\pm$25.2 \\
5434.524& \fei&   1.011 &  --2.122 &189.0$\pm$22.1 &170.1$\pm$20.4 &167.7$\pm$20.5 &178.7$\pm$24.4 &203.9$\pm$24.5 &180.3$\pm$21.3 \\
5446.916& \fei&   0.990 &  --1.914 &200.5$\pm$22.6 &199.9$\pm$23.0 &174.5$\pm$21.0 &\dots &261.3$\pm$28.4 &217.8$\pm$24.1 \\
5455.609& \fei&   1.011 &  --2.091 &225.6$\pm$25.2 &198.8$\pm$23.0 &188.3$\pm$21.9 &191.9$\pm$24.5 &286.9$\pm$30.3 &244.5$\pm$26.2 \\
5501.465& \fei&   0.958 &  --3.047 &135.3$\pm$15.7 &115.1$\pm$13.9 &103.4$\pm$13.7 &106.4$\pm$15.4 &170.4$\pm$19.0 &142.1$\pm$16.2 \\
5506.779& \fei&   0.990 &  --2.797 &149.4$\pm$17.1 &123.4$\pm$14.7 &129.6$\pm$16.1 &133.9$\pm$18.1 &175.9$\pm$19.8 &148.3$\pm$16.9 \\
5528.405& \mgi&   4.346 &  --0.620 &117.5$\pm$14.4 &95.7$\pm$12.1 &100.5$\pm$13.0 &116.2$\pm$16.0 &147.4$\pm$16.5 &136.7$\pm$15.7 \\
5581.965& \cai&   2.523 &  --0.555 &41.9$\pm$7.1 &\dots &56.0$\pm$9.5 &62.1$\pm$11.6 &82.0$\pm$10.1 &66.4$\pm$8.7 \\

\hline
\multicolumn{10}{c}{\bf Giraffe HR13}\\
6122.217& \cai&   1.886 &  --0.316  &110.7$\pm$12.7 &111.1$\pm$13.3 &81.4$\pm$10.5 &\dots &155.7$\pm$16.9 &137.0$\pm$15.4 \\
6137.691& \fei&   2.588 &  --1.403  &112.9$\pm$13.3 &101.6$\pm$13.3 &89.5$\pm$11.8 &102.4$\pm$16.8 &139.7$\pm$15.5 &120.5$\pm$13.9 \\
6161.297& \cai&   2.523 &  --1.266  &\dots &\dots &18.1$\pm$4.2 &\dots &41.8$\pm$5.7 &30.9$\pm$5.0 \\
6162.173& \cai&   1.899 &  --0.090  &126.5$\pm$14.6 &108.4$\pm$13.3 &102.5$\pm$12.9 &\dots &169.2$\pm$20.0 &154.7$\pm$18.1 \\
6166.439& \cai&   2.521 &  --1.142  &\dots &\dots &\dots &\dots &50.8$\pm$9.8 &42.8$\pm$7.4 \\
6213.430& \fei&   2.223 &  --2.482  &62.7$\pm$8.0 &63.4$\pm$8.7 &45.6$\pm$7.0 &73.7$\pm$11.5 &111.2$\pm$12.5 &89.1$\pm$10.5 \\
6219.281& \fei&   2.198 &  --2.433  &72.9$\pm$9.2 &71.5$\pm$9.7 &54.8$\pm$8.2 &53.9$\pm$9.4 &114.8$\pm$13.0 &101.8$\pm$11.8 \\
6230.722& \fei&   2.559 &  --1.281  &108.5$\pm$12.9 &135.4$\pm$16.7 &95.3$\pm$12.4 &102.2$\pm$14.8 &162.0$\pm$17.9 &143.1$\pm$16.7 \\
6252.555& \fei&   2.404 &  --1.687  &109.0$\pm$13.0 &96.6$\pm$12.2 &85.3$\pm$11.5 &109.7$\pm$16.2 &143.0$\pm$16.5 &121.0$\pm$14.2 \\
6254.258& \fei&   2.279 &  --2.443  &71.0$\pm$9.5 &\dots &44.0$\pm$7.7 &\dots &117.3$\pm$14.2 &101.8$\pm$12.8 \\
6256.361& \fei&   2.453 &  --2.408  &62.0$\pm$8.8 &56.4$\pm$8.1 &\dots &\dots &107.3$\pm$14.0 &91.2$\pm$12.8 \\
6270.223& \fei&   2.858 &  --2.464  &15.7$\pm$3.2 &\dots &\dots &\dots &52.6$\pm$6.8 &36.3$\pm$5.5 \\
6318.018& \fei&   2.453 &  --2.261  &85.5$\pm$11.3 &86.7$\pm$11.8 &56.6$\pm$8.5 &70.4$\pm$11.2 &121.5$\pm$14.1 &100.6$\pm$12.4 \\
6322.685& \fei&   2.588 &  --2.426  &42.7$\pm$6.2 &46.1$\pm$7.5 &15.5$\pm$3.9 &\dots &83.1$\pm$9.7 &56.9$\pm$7.5 \\
6335.330& \fei&   2.198 &  --2.177  &86.4$\pm$10.5 &\dots &\dots &100.3$\pm$15.8 &126.3$\pm$14.0 &112.7$\pm$13.2 \\
6336.823& \fei&   3.686 &  --0.856  &47.3$\pm$6.7 &49.8$\pm$8.0 &51.2$\pm$8.3 &69.4$\pm$13.1 &84.5$\pm$10.2 &73.4$\pm$9.3 \\

\hline
\multicolumn{10}{c}{\bf Giraffe HR14}\\
6393.600& \fei&   2.433 &  --1.432 &100.0$\pm$12.0 &124.2$\pm$15.3 &96.2$\pm$12.8 &\dots &146.4$\pm$16.2 &135.9$\pm$15.5 \\
6411.648& \fei&   3.654 &  --0.595 &57.3$\pm$7.6 &65.8$\pm$9.5 &70.4$\pm$10.3 &96.0$\pm$15.9 &108.6$\pm$12.3 &90.3$\pm$10.8 \\
6430.845& \fei&   2.176 &  --2.006 &112.1$\pm$13.3 &99.7$\pm$12.5 &97.6$\pm$13.6 &84.0$\pm$12.9 &150.0$\pm$16.4 &133.6$\pm$15.3 \\
6439.075& \cai&   2.526 &   0.390 &89.6$\pm$11.0 &95.7$\pm$11.8 &\dots &102.4$\pm$14.9 &148.0$\pm$16.2 &113.8$\pm$12.9 \\
6494.980& \fei&   2.404 &  --1.273 &122.1$\pm$15.7 &128.8$\pm$16.9 &105.4$\pm$14.8 &125.7$\pm$19.3 &165.8$\pm$21.8 &137.4$\pm$19.5 \\
6592.913& \fei&   2.727 &  --1.473 &79.6$\pm$9.4 &79.6$\pm$9.8 &57.0$\pm$8.2 &80.6$\pm$12.8 &132.1$\pm$14.4 &103.1$\pm$11.7 \\
6593.870& \fei&   2.433 &  --2.422 &65.4$\pm$8.1 &\dots &43.3$\pm$7.0 &\dots &106.7$\pm$11.9 &79.6$\pm$9.4 \\

\hline

\end{tabular}
\tablefoot{This table is available entirety in a machine-readable form in the 
online journal.}
\end{table*}
\begin{table*}
\footnotesize
\caption{Individual abundances ($\log \epsilon$) and errors for old population stacked stars in Giraffe LR08 grism.}
\label{tab_LRold}
\centering
\begin{tabular}{cccc cccccc }
\hline\hline
 {$\lambda$ (\AA)}&
 {Elem.}&
 {$\chi$ (eV)}&
 {$\log gf$}&
 {LRold1}&
 {LRold2}&
 {LRold3}&
 {LRold4}&
 {LRold5}&
 {LRold6\dots}\\
 \hline
8327.056&   \fei&   2.198 &  --1.525 &5.32$\pm$0.13 & 5.50$\pm$0.14 & 5.50$\pm$0.14 & 5.69$\pm$0.14 & 5.57$\pm$0.14 & 5.58$\pm$0.14 \\ 
8387.772&   \fei&   2.176 &  --1.493 &4.90$\pm$0.12 & 5.30$\pm$0.13 & 5.50$\pm$0.14 & 5.63$\pm$0.14 & 5.49$\pm$0.14 & 5.47$\pm$0.14 \\ 
8688.624&   \fei&   2.176 &  --1.212 &4.91$\pm$0.12 & 5.31$\pm$0.13 & 5.50$\pm$0.14 & 5.49$\pm$0.14 & 5.49$\pm$0.14 & 5.61$\pm$0.14 \\ 
8806.756&   \mgi&   4.346 &  --0.137 &5.56$\pm$0.14 & 5.70$\pm$0.14 & 5.69$\pm$0.14 & 5.71$\pm$0.14 & 5.59$\pm$0.14 & 5.70$\pm$0.14 \\ 

\hline

\end{tabular}
\tablefoot{This table is available entirety in a machine-readable form in the 
online journal.}
\end{table*}

\begin{table*}
\footnotesize
\caption{Individual abundances ($\log\epsilon$) and errors for intermediate-age population stacked stars in Giraffe LR08 grism}
\label{tab_LRint}
\centering
\begin{tabular}{cccc cccccc}
\hline\hline
 {$\lambda$ (\AA)}&
 {Elem.}&
 {$\chi$ (eV)}&
 {$\log gf$}&
 {LRint1}&
 {LRint2}&
 {LRint3}&
 {LRint4}&
 {LRint5}&
 {LRint6\dots}\\
\hline
8327.056&   \fei&   2.198 &  --1.525 &5.79$\pm$0.14 & 5.88$\pm$0.15 & 5.94$\pm$0.15 & 6.09$\pm$0.15 & 6.20$\pm$0.15 & 6.08$\pm$0.15 \\ 
8387.772&   \fei&   2.176 &  --1.493 &5.50$\pm$0.14 & 5.51$\pm$0.14 & 5.65$\pm$0.14 & 5.52$\pm$0.14 & 5.88$\pm$0.15 & 5.90$\pm$0.15 \\ 
8688.624&   \fei&   2.176 &  --1.212 &5.51$\pm$0.14 & 5.66$\pm$0.14 & 5.52$\pm$0.14 & 5.70$\pm$0.14 & 5.90$\pm$0.15 & 6.10$\pm$0.15 \\ 
8806.756&   \mgi&   4.346 &  --0.137 &6.16$\pm$0.15 & 6.30$\pm$0.16 & 6.16$\pm$0.15 & 6.15$\pm$0.15 & 6.16$\pm$0.15 & 6.36$\pm$0.16 \\ 

\hline

\end{tabular}
\tablefoot{This table is available entirety in a machine-readable form in the 
online journal.}
\end{table*}

        \subsection{UVES equivalent widths}
        \label{ew_meas}

The elemental abundances for the UVES and FLAMES/UVES
spectra were determined from equivalent
width (EW) measurements. EWs were measured with a proprietary
IDL\footnote{IDL is distributed by the Exelis Visual Information
Solutions.} interactive procedure, based on a Gaussian or Voigt fitting
routine. The user controls the continuum placement, the
profile of individual lines, and the contribution of the wings to the EW
values. Continuum estimation, in particular, is crucial for
the robustness of the final results. 
To minimize any systematic bias in the 
continuum estimate that is due to the subjectivity of the operator, three of 
us have independently performed EW measurements on a sample of 
selected lines (weak and strong, high and low S/N). The
internal dispersion is lower than 6~m\AA,\ and there is no evidence of
systematics. We also performed a sanity check on the profile measurement 
using the IRAF\footnote{IRAF is distributed by the National Optical Astronomy
Observatory, which is operated by the Association of Universities for
Research in Astronomy, Inc., under cooperative agreement with the
National Science Foundation.} task {\tt splot}. The differences 
are within few percent.

We estimated the uncertainties in the equivalent widths
(EW$_{\rm rms}$) using the formula presented by \citet{cayrel88},
revisited by \citet{venn12}:
$$
{\rm EW_{rms}}=({\rm S/N})^{-1}\times\sqrt{1.5\times{\rm
FWHM}\times\delta x,}
$$
where S/N is the signal-to-noise ratio per pixel, FWHM is the line full
width at half-maximum, and $\delta x$ is the pixel size. Following this
approach, we adopted a more conservative EW error:
$$
\epsilon{\rm EW}={\rm EW_{rms}}+0.1\times{\rm EW}.
$$
This conservative approach, which we consider robust, gives
EW$_{\rm rms}$$\approx$2~m\AA\ for the whole sample with a final error
$\epsilon{\rm EW}$$\approx$10~m\AA. Measured EWs with errors are listed 
in Table~\ref{tab_UVES}.

To evaluate the precision of our EWs, we compared the measurements of
non-iron group lines (listed in Table~\ref{tab_UVES}) with those
available in the literature.
Specifically, we compared EWs from \citet{shetrone03}, \citet{koch08}, and
\citet{venn12}, based on UVES and on \giraffe-UVES spectra,
and from \citet{lemasle12}, based on \giraffe-HR spectra.
Figure~\ref{ew} shows the EW comparison for the four samples, with 
our measurements always on the x-axis. The
top left panel represents the sample of \citet{shetrone03}, with which we 
have five stars in common (one symbol per stars).
The black dashed line represents equality, and the dotted lines show
a 10\%\ error convolved with the 10~m\AA\ error, following \citet{shetrone03}.
The error bars in the right bottom corner 
display the mean errors of the two EW measurements. The mean difference
and the dispersion are also labeled (in unit of m\AA). The comparison
shows that our estimates are higher on average by $\sim$8~m\AA, but
the measurements agree well within 10\%. We attribute
these systematic differences to the continuum normalization,
since a typical uncertainty of 10\%\ on the location of the
continuum causes a difference of 10\%\ in the EW. 

The top right panel of Fig.~\ref{ew} shows the same
comparison for the sample of \citet{koch08} (ten stars). The higher dispersion ($\sim$20~m\AA) is
mainly due to the low S/N of these spectra, while the mean
difference is about 12~m\AA. Once again, we
overestimated the EWs. The bottom left panel shows the
comparison for eleven stars in common with the sample of \citet{lemasle12}. 
In this case, the systematic difference is larger ($\sim$--19~m\AA), 
but here our EW estimates are lower. The high dispersion 
seems to be caused by the different spectral resolution (GHR$\sim$20,000 vs.
UVES$\sim$40,000), and the mean error on EWs decreases by almost 
a factor of two (20.5 vs $\sim$12~m\AA).
The bottom right panel shows the comparison
with the recent work of \citet{venn12} (seven stars, six of them
reanalyzed by us). In this case, we obtain a difference of 
$\sim$--10~m\AA\ with a dispersion of 18~m\AA, which is mainly due to the modest
S/N of these spectra (10-30). 

In conclusion, the data plotted in Fig.~\ref{ew} indicate that
the current EWs agree on average with similar estimates
available in the literature, within 10-15\%. 

        \section{Abundances}
\label{sec:abund}
        \subsection{Model atmospheres}
The individual model atmospheres come from the interpolation on the \marcs\
grid \citep{gustafsson08}, using a modified version of the interpolation
code developed by \citet{masseron06}. The individual models were computed 
for the stellar parameters ($\teff, \log g$) listed in Table~1
of Paper~V and in Table~\ref{tab_param_stack}.
Moreover, we selected models with
spherical geometry, an $\alpha$-enhanced (\afe=+0.4)
chemical mixture, a mass value of 1~\msun\ , and a constant
microturbulence velocity ($\xi$=2~\kms), as described in Paper~V. It is
noteworthy that we did not include lines shortward of 4800~\AA\ 
to avoid any possible continuum scattering effect in this
wavelength region \citep{sobeck11}. 

        \subsection{UVES abundances}
        \label{ew_abund}        
For the abundance determinations, we used the 2010 version of the
stellar abundance code \moog\ 
\citep{sneden73}\footnote{\tt http://www.as.utexas.edu/$\sim$chris/moog.html}, 
in particular its {\it abfind} driver. 
The abundances presented in the following sub-sections were computed
with a 1D LTE analysis. We chose the solar chemical
composition from \citet{grevesse07} to be consistent with the iron
abundances derived in Paper~V. The reference values adopted for the
individual species and abundance results are listed in Table~\ref{tab_abund}. 
The anonymous referee suggested to provide more quantitative estimates
concerning the upper limits on the abundances of weak lines (O, Na, Si)
in metal-poor stars. To constrain the above limits, we performed a series
of simulations using synthetic and observed spectra with S/N$\ge$40. We
found that we can measure lines with EWs larger than 11~m\AA\ for stars
with iron abundances ranging from \feh=--1.50 to \feh=--2.50. The quoted
limit implies upper limits in the abundance of the quoted elements of
about \nafe=--0.9$\div$0.2, \ofe=--0.1$\div$0.6 and
\sife=--0.6$\div$0.4. We performed the same test using spectra with
lower S/N and found that we can only measure lines with EWs larger
than 20~m\AA. This means upper limits in the abundances of
\nafe=--0.3$\div$0.8, \ofe=0.2$\div$0.9 and \sife=--0.3$\div$0.7.
        \subsubsection{Comparison with literature values}
\begin{figure}
\centering
\includegraphics[trim=0.5cm 0cm 4cm 0.5cm,clip,width=0.5\textwidth]{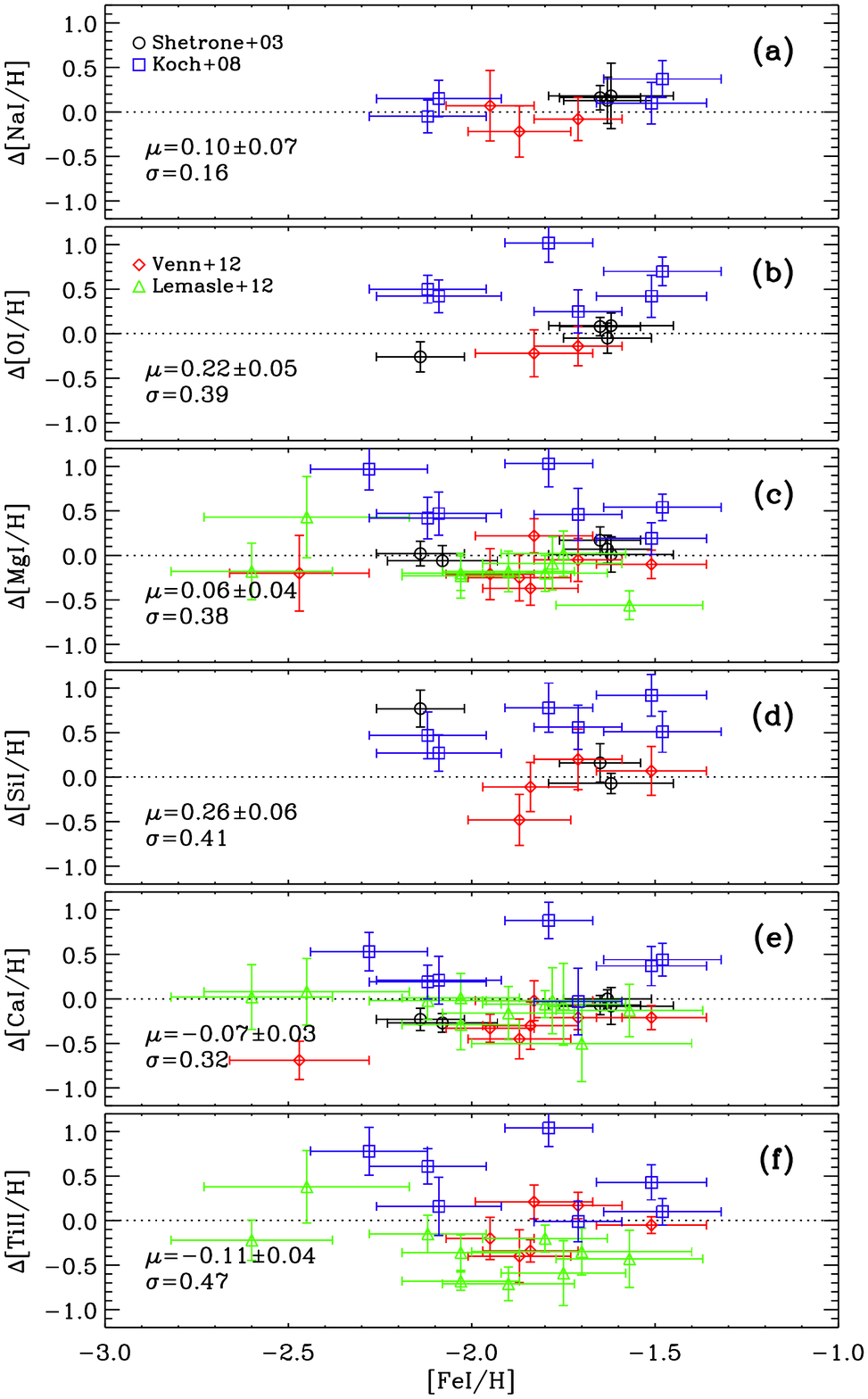}
\caption{Comparison of UVES abundances with the literature
data samples indicated, $\Delta$\xh=\xh$_{\rm UVES}$--\xh$_{\rm Other}$.
\label{abcompare}} 
\end{figure}

Figure~\ref{abcompare} shows the individual UVES abundance results
obtained in this work compared to literature values (rescaled to the
same solar reference abundances). In particular, each panel of
Fig.~\ref{abcompare} shows the $\Delta\xh$=\xh$_{\rm UVES}$--\xh$_{\rm
Other}$ as a function of \feh\  for the stars in common with
\citet[][black circles]{shetrone03}, \citet[][blue
squares]{koch08}, \citet[][red diamonds]{venn12}, and
\citet[][green triangles]{lemasle12}.
The error bars plotted in this figure were estimated by summing in
quadrature current uncertainties with uncertainties evaluated by the quoted
authors.
The current abundances agree, within 1$\sigma$, with 
high-resolution abundances available in the literature, namely 
\citet{shetrone03}, \citet{venn12}, and \citet{lemasle12}. 
The abundances by \citet{koch08} show a systematic offset and a large 
scatter when compared with our measurements. The quoted discrepancy 
appears to be caused by the differences in the measured EWs 
(see Sect. \ref{ew_meas} and also Fig.~\ref{ew}) and in the adopted 
stellar parameters. Their surface gravities are higher on average by 
0.5~dex than current ones. The difference seems to be due to 
the different approach adopted to estimate the gravity, that
is, by forcing
the balance between \fei\ and \feii\ vs. photometric gravities. 
A more detailed discussion is reported in Sect. 5.2 of Paper~V. 
Owing to the lack of evident trends and significant 
systematics with the estimates available in the literature, we did not 
apply any correction to our UVES abundances.

        \subsection{\giraffe-HR abundances}
\begin{figure}
\centering
\includegraphics[trim=0.cm 6.5cm 1cm 0.5cm,clip,width=0.95\columnwidth]{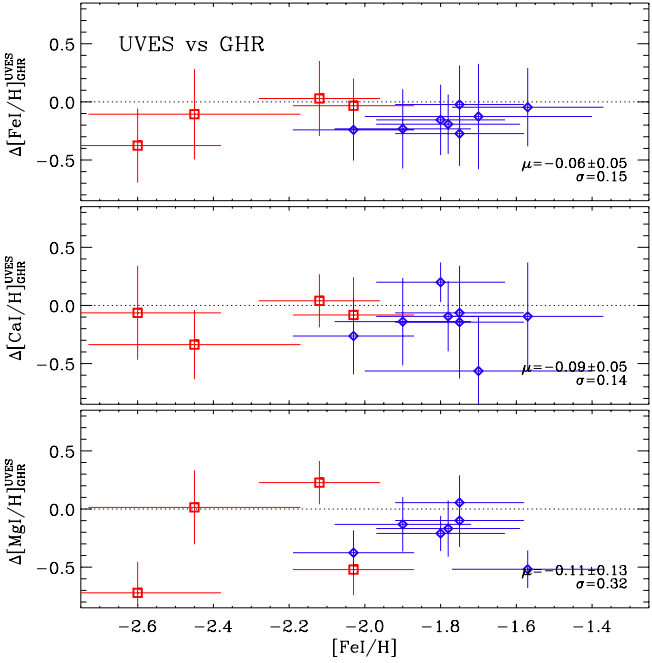}
\caption{Comparison of \fei, \cai,\ and \mgi\ abundances between
UVES and individual Giraffe HR spectra, 
$\Delta$\xh=\xh$_{\rm UVES}$--\xh$_{\rm GHR}$. 
Red squares and blue diamonds show abundances of old and intermediate-age stars.
\label{uves-hr}} 
\end{figure}

The approach described in Sects. \ref{ew_meas} and
\ref{ew_abund} was also used to measure the EWs (see Table~\ref{tab_HR})
and obtain
chemical abundances for the stacked \giraffe-HR spectra 
(see Table~\ref{tab_abund}). 
To check the validity of our measurements on the stacked spectra
and to avoid any systematics, we performed the same analysis
on the individual spectra for the stars in common with the UVES
sample. We compare the abundances of \fei, \cai\ and \mgi\ as
functions of \feh\ in Fig.~\ref{uves-hr}. 
The agreement is good, within 1$\sigma$ (see labeled values), for most 
measurements and without evidence of a drift as a function of \feh.

The bottom panel of Fig.~\ref{uves-hr} shows that four objects display a
difference in \mgi\ abundance that is larger than 1$\sigma$. In
particular, the difference for the most metal-poor (Car40) and the most
metal-rich (Car51) is about 2$\sigma$. We double-checked these objects 
together with Car27 and Car33, located at \feh$\approx$--2,
and we found that they are the faintest targets in the UVES data sample,
meaning they have the lowest signal-to-noise ratio.
Moreover, the continuum in the region bracketing the only available Mg line
($\approx$5528~\AA) is relatively noisy. The EWs based on UVES data
show a mean difference of $\sim$50~m\AA\ with those based on HR spectra. 
We note that these differences do not affect the results of
this investigation.

        \subsection{Abundance uncertainties}
        \label{subsec:ab_uncert}
The abundance errors were estimated as the maximum of two
values. The first comes from propagation 
of the errors in the EW measurements
($\epsilon$EW), estimated following the approach described in
Sect.~\ref{ew_meas}, to obtain a $\sigma$(EW) for each
line. When the quoted error was asymmetric, the
average value was adopted. The second error value was based on the
standard deviation of the abundances if more than three lines of the
element were available $\sigma(X)$. Otherwise, we set
$\sigma(X)$=$\sigma(\fei)$. Moreover, to account for the
uncertainties in the stellar parameters, we added in quadrature
the contributions coming from the following error budget:
we computed the abundance variations by changing,
one at a time, the temperature ($\pm$75~K), gravity
($\pm$0.2~dex), microturbulence ($\pm$0.25~\kms), equivalent width
($\pm$10~m\AA), $\log gf$ ($\pm$0.15), metallicity ($\pm$0.2~dex),
and $\alpha$-content ($\pm$0.4~dex). We note that we used
generous estimates for the uncertainties in the atmospheric
parameters (see Paper~V) to include the differences
between our set of parameters, models, and atomic data as compared
to the literature ones. The estimation was performed on the star
Car12, since its effective temperature ($\sim$4400~K) and surface
gravity ($\sim$0.80~dex) can be considered representative of
the entire sample. The results are listed in Table~\ref{tab_err}. 
For the \giraffe-HR stacked spectra, the dispersion of
individual spectra (see the top panel of Fig.~\ref{specstack}) produces 
an uncertainty in the measured EWs of about 10\%. In terms of 
abundances, this effect results in an uncertainty of $\sim$0.15~dex.

\begin{table*}
\footnotesize
\caption{Impact of uncertainties on abundances for the representative star Car12.}
\label{tab_err}
\centering
\begin{tabular}{lcccccccccccccc}
\hline\hline
 &
 \multicolumn{2}{c}{$\Delta\teff$}&
 \multicolumn{2}{c}{$\Delta\log g$}&
 \multicolumn{2}{c}{$\Delta\xi$}&
 \multicolumn{2}{c}{$\Delta$EW}&
 \multicolumn{2}{c}{$\Delta\log gf$}&
 \multicolumn{2}{c}{$\Delta\feh$}&
 \multicolumn{1}{c}{$\Delta\afe$}\\
 &
 \multicolumn{2}{c}{(K)}&
 \multicolumn{2}{c}{(dex)}&
 \multicolumn{2}{c}{(\kms)}&
 \multicolumn{2}{c}{(m\AA)}&
 \multicolumn{2}{c}{}&
 \multicolumn{2}{c}{(dex)}&
 \multicolumn{1}{c}{(dex)}\\
 {Elem.}&
 {$-75$}&
 {$+75$}&
 {$-0.2$}&
 {$+0.2$}&
 {$-0.25$}&
 {$+0.25$}&
 {$-10$}&
 {$+10$}&
 {$-0.15$}&
 {$+0.15$}&
 {$-0.2$}&
 {$+0.2$}&
 {$-0.4$}&
 {$<$$\sigma$$>$\tablefootmark{a}}\\
\hline

\nai  &  $-0.07$  &  $+0.07$  &  $+0.01$  &  $+0.00$  &  $+0.01$  &  $-0.01$  &  $-0.30$  &  $+0.00$  &  $+0.15$  &  $-0.15$  &  $+0.02$  &  $-0.01$  &  $+0.04$  &  $+0.27$\\
\oi  &  $-0.02$  &  $+0.02$  &  $-0.08$  &  $+0.09$  &  $+0.01$  &  $-0.01$  &  $-0.21$  &  $+0.05$  &  $+0.15$  &  $-0.15$  &  $-0.06$  &  $+0.07$  &  $-0.09$  &  $+0.25$\\
\mgi  &  $-0.06$  &  $+0.07$  &  $+0.02$  &  $-0.01$  &  $+0.10$  &  $-0.09$  &  $-0.25$  &  $+0.03$  &  $+0.15$  &  $-0.14$  &  $+0.02$  &  $-0.01$  &  $+0.03$  &  $+0.26$\\
\sii  &  $+0.02$  &  $+0.01$  &  $-0.01$  &  $+0.03$  &  $+0.02$  &  $-0.01$  &  $-0.32$  &  $+0.11$  &  $+0.15$  &  $-0.15$  &  $-0.01$  &  $+0.02$  &  $-0.01$  &  $+0.28$\\
\cai  &  $-0.10$  &  $+0.10$  &  $+0.00$  &  $+0.01$  &  $+0.12$  &  $-0.10$  &  $-0.14$  &  $+0.18$  &  $+0.15$  &  $-0.15$  &  $+0.04$  &  $-0.02$  &  $+0.04$  &  $+0.27$\\
\tiii  &  $+0.02$  &  $-0.02$  &  $-0.07$  &  $+0.07$  &  $+0.16$  &  $-0.14$  &  $-0.18$  &  $+0.20$  &  $+0.15$  &  $-0.15$  &  $-0.04$  &  $+0.04$  &  $-0.06$  &  $+0.30$\\
\fei  &  $-0.07$  &  $+0.09$  &  $-0.02$  &  $+0.02$  &  $+0.05$  &  $-0.04$  &  $-0.17$  &  $+0.15$  &  $+0.15$  &  $-0.15$  &  $+0.01$  &  $+0.00$  &  $+0.02$  &  $+0.24$\\
\feii  &  $+0.08$  &  $-0.06$  &  $-0.08$  &  $+0.08$  &  $+0.07$  &  $-0.06$  &  $-0.18$  &  $+0.18$  &  $+0.15$  &  $-0.15$  &  $-0.06$  &  $+0.06$  &  $-0.09$  &  $+0.28$\\
\hline
\end{tabular}
\tablefoot{
\tablefoottext{a}{Weighted standard deviation.}\\
}

\end{table*}

        \subsection{\giraffe-MR abundance}
\begin{figure}
\centering
\includegraphics[trim=0cm 13cm 1cm 0.5cm,clip,width=0.95\columnwidth]{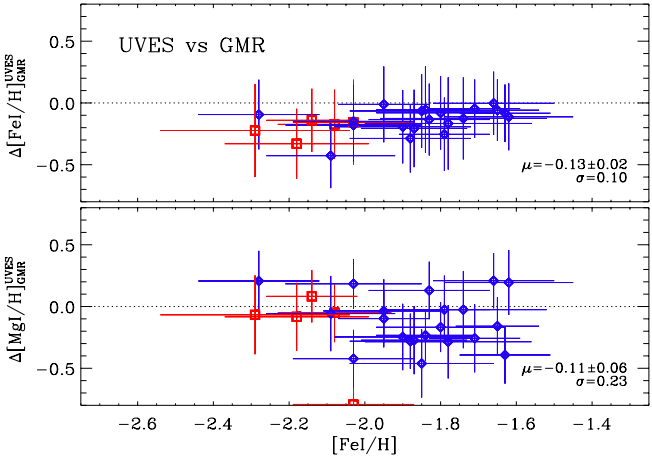}
\caption{Comparison of \fei and \mgi\ abundances between UVES and
individual Giraffe MR spectra, 
$\Delta$\xh=\xh$_{\rm UVES}$--\xh$_{\rm GMR}$ 
\label{uves-lr}} 
\end{figure}

\begin{figure}
\centering
\includegraphics[trim=0cm 13cm 1cm 0.5cm,clip,width=0.95\columnwidth]{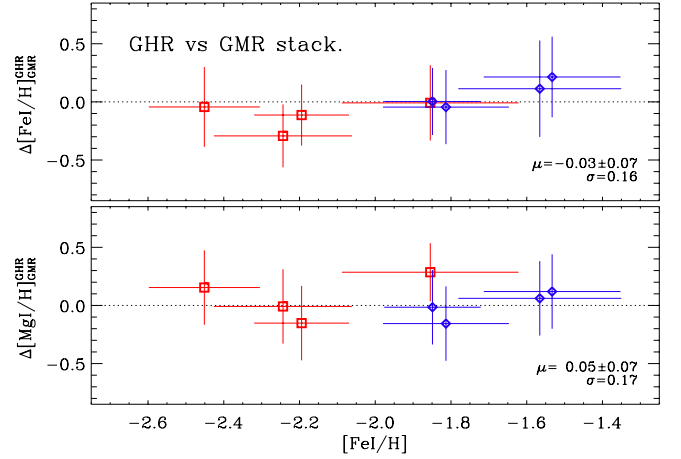}
\caption{Comparison of \fei\ and \mgi\ abundances between stacked
Giraffe HR and LR08 spectra, 
$\Delta$\xh=\xh$_{\rm GHR}$--\xh$_{\rm GMR}$ 
\label{hr-lr}}
\end{figure}

The spectral features in the \giraffe-MR data are severely
affected by the blending effect that is caused by the medium resolution of
the spectra (R$\sim$6,000). Equivalent width measurements are
thus not reliable; to distinguish the contribution of the various blends, 
synthetic spectra need to be computed. For this, we used 
the \textit{synth} driver of \moog. The synthetic spectra were
convolved with a Gaussian broadening function to reproduce the low
instrumental resolution. We excluded the effect of stellar rotation.
The synthetic spectra were computed for various
abundances of iron, magnesium, and calcium. Then they
were compared, line by line, with the observed spectra. The
resulting abundance for each line was measured from the
minimum of the residual function. 
The uncertainties for individual lines were estimated as the
sum in quadrature of three contributions: the abundance step adopted in
spectral synthesis computations, the error in the quadratic fit used to
interpolate the residual function, and the resulting uncertainty in the
abundances ($\sim$0.15~dex) that is due to the dispersion of individual spectra
(see bottom panel of Fig.~\ref{specstack} and Sect. \ref{subsec:ab_uncert}).
Measured abundances with 
errors are listed in Tables~\ref{tab_LRold} and \ref{tab_LRint}.
To verify the validity
of our measurements on the stacked spectra and to avoid any
systematics, we performed the same analysis on the individual
spectra for the stars in common with the UVES sample. We
compare the abundances of \fei\ and \mgi\ as function of \feh\ in
Fig.~\ref{uves-lr}. 
The agreement is good, within 1$\sigma$ (see labeled values), for most 
measurements and without evidence of a trend as a function of \feh.
Figure~\ref{hr-lr} shows the comparison between the resulting
abundances of \fei\ and \mgi\ from stacked \giraffe-HR and -MR
spectra. We do not find any significant systematic trends
between the two data sets. 
We note that the two objects that in the bottom panel display a difference  
of about 2$\sigma$ are once again Car27 and Car33, that is, the faintest
tail of UVES targets.

\section{Abundances of old and intermediate-age stars}
\label{sec:abund_vs_logg}
\begin{figure*}[!ht]
\centering
\includegraphics[trim=0cm 9cm 1cm 0.5cm,clip,width=0.7\textwidth]{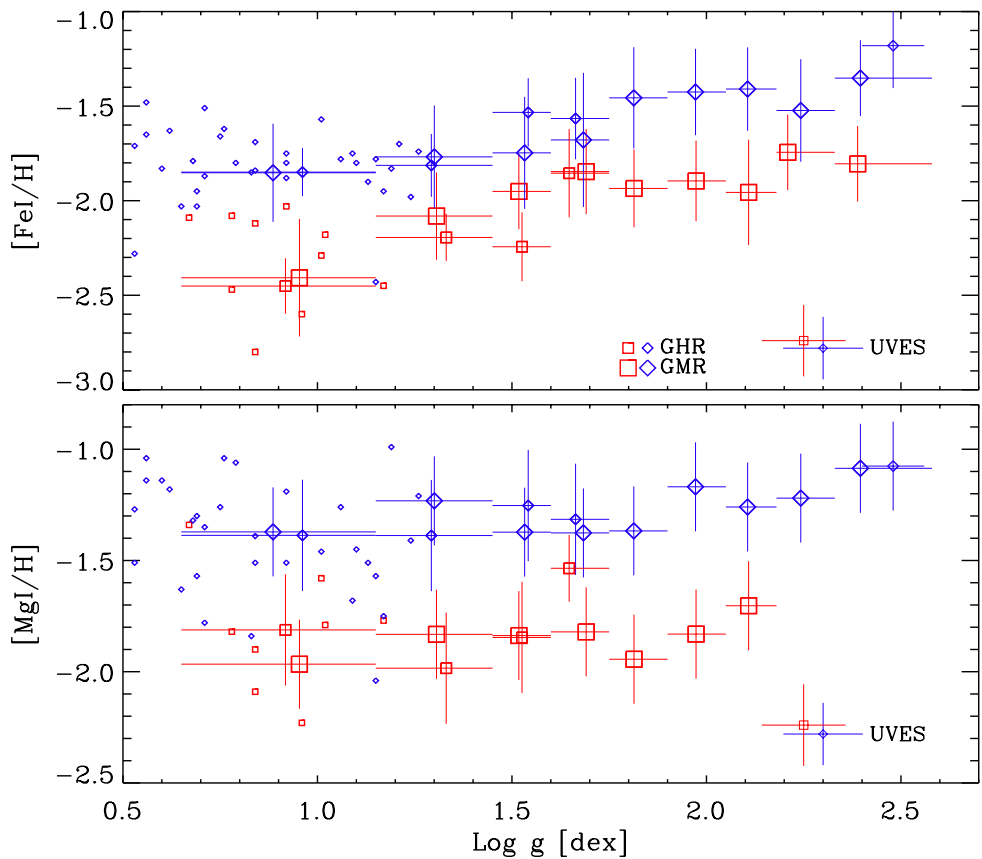}
\caption{Top: \feh\ abundances based on individual and stacked spectra. Red squares 
and blue diamonds represent abundances of old and intermediate-age stars.
Abundances based on individual high-resolution UVES spectra are
displayed as small squares and diamonds, without bars. 
The error bars plotted in the bottom right
corner of the panel display the typical uncertainty for the UVES
abundances and on surface gravities (see also Paper~V).
Abundances based
on GHR spectra are marked by medium squares and diamonds, 
while those based on GMR spectra are marked by large squares/diamonds. 
The vertical bars represent the uncertainty in
iron while the horizontal ones show the gravity ranges adopted in
Fig.~\ref{cmdstack}.
Bottom: same as the top, but for the \mgh\ abundances.
\label{ab-logg}}
\end{figure*}

The resulting abundances for individual and stacked spectra are
listed in Table~\ref{tab_abund}. Figure~\ref{ab-logg} shows the
\fei\ and \mgi\ abundances as function of gravity for the whole
data set. As usual, the red squares are used for the
old and the blue diamonds for the intermediate-age
population. The plots show an evident dichotomy in the abundances that
covers the entire gravity range, from the top of the RGB (\logg$\simeq$0.5~dex) 
to the RC level ($\sim$2.5~dex). 

This figure presents several interesting features. \\ 
{\it (i)} Iron abundances (top panel) based on UVES, GHR, and GMR
spectra show that the old stellar population is, over the entire
gravity range, systematically more metal-poor than the
intermediate-age stellar population. The mean iron abundances
based on the three different sets of spectra are listed in
Table~\ref{tab_cl}. The weighted total mean for the old population
is \feh=--2.15$\pm$0.06 ($\sigma$=0.28), while for the
intermediate-age population it is \feh=--1.75$\pm$0.03
($\sigma$=0.21). The difference is slightly larger than 1$\sigma$. 
To provide a more quantitative estimate, we smoothed the
metallicity distributions of the old and intermediate-age
data sets with a Gaussian kernel with unitary weight and sigma
equal to the individual abundance uncertainties. We performed a
$\chi^2$ comparison of the two distributions and the confidence levels (CL)
are listed in Col. 4 of Table~\ref{tab_cl}. These data
indicate that the iron abundances of the two stellar
populations differ with a confidence level that ranges from 
75\%\ (global sample) to 84\%\ (GHR).\\
{\it (ii)} Magnesium abundances plotted in the bottom panel of 
Fig.~\ref{ab-logg}
display a similar 
trend. The mean abundances for the different
spectroscopic samples are also listed in Table~\ref{tab_cl}. The
mean magnesium abundance for the old population based on the
entire sample is \mgh=--1.91$\pm$0.05 ($\sigma$=0.22), while for
the intermediate-age population it is \mgh=--1.35$\pm$0.03
($\sigma$=0.22). The difference is slightly larger than
1$\sigma$. We followed the same approach adopted for the iron
abundances and found that they differ with a confidence level
that ranges from 80\%\ (GHR) to 91\%\ (GMR).\\ 
{\it (iii)} The iron and the magnesium abundances based on GHR and
GMR spectra agree in the overlapping surface gravity regime,
with individual abundances based on UVES spectra.\\
{\it (iv)} The largest surface gravity bin (\logg=2.48) shows
the Fe and the Mg abundances of RC stars. The abundances are,
within the errors, similar to the other intermediate-age
abundances. This further confirms the difference between the
two subpopulations, since RC stars are reliable tracers of the
intermediate-age population \citep{cassisi13book}. 

\begin{table*}
\footnotesize
\caption{Mean chemical abundances and dispersions of Carina stars.}
\label{tab_abund}
\centering
\begin{tabular}{l cccccccc}
\hline\hline
 {ID}&
 {\oife}&
 {\naife}&
 {\mgife}&
 {\siife}&
 {\caife}&
 {\tiiife}&
 {\feih}\\
 
\hline
Solar Ref. & 8.66 & 6.17 & 7.53 & 7.51 & 6.31 & 4.90 & 7.45 \\
\hline
\multicolumn{8}{c}{\bf OLD} \\
Car3 &0.51$\pm$0.12[1]&\dots &0.36$\pm$0.07[2]&1.09$\pm$0.12[1]&0.28$\pm$0.11[10]&0.87$\pm$0.18[7]&--2.14$\pm$0.12[38]\\
Car10 &\dots &\dots &0.26$\pm$0.13[2]&\dots &--0.03$\pm$0.09[10]&0.39$\pm$0.19[5]&--2.08$\pm$0.15[12]\\
Car13 &\dots &\dots &0.71$\pm$0.19[1]&\dots &0.86$\pm$0.06[11]&0.66$\pm$0.12[4]&--2.80$\pm$0.19[19]\\
Car27 &\dots &\dots &--0.54$\pm$0.16[1]&\dots &--0.33$\pm$0.27[8]&--0.08$\pm$0.19[5]&--2.03$\pm$0.16[9]\\
Car30 &\dots &\dots &0.39$\pm$0.19[1]&\dots &0.14$\pm$0.44[11]&--0.12$\pm$0.26[4]&--2.18$\pm$0.19[9]\\

\hline
\multicolumn{8}{c}{\bf INTERMEDIATE}\\
Car2 &0.66$\pm$0.12[1]&0.01$\pm$0.24[3]&0.45$\pm$0.12[1]&\dots &0.35$\pm$0.09[13]&0.42$\pm$0.13[7]&--1.63$\pm$0.12[17]\\
Car12 &0.71$\pm$0.10[2]&0.36$\pm$0.35[2]&0.58$\pm$0.17[2]&0.47$\pm$0.09[3]&0.37$\pm$0.20[11]&0.58$\pm$0.14[6]&--1.62$\pm$0.17[21]\\
Car14 &0.46$\pm$0.12[1]&\dots &0.47$\pm$0.19[2]&0.22$\pm$0.21[2]&0.32$\pm$0.10[13]&0.48$\pm$0.11[6]&--1.79$\pm$0.12[25]\\
Car15 &\dots &\dots &0.77$\pm$0.14[2]&\dots &0.33$\pm$0.10[11]&0.58$\pm$0.19[5]&--2.28$\pm$0.16[27]\\
Car16 &0.52$\pm$0.17[1]&0.33$\pm$0.17[1]&0.56$\pm$0.18[2]&\dots &0.36$\pm$0.45[14]&0.22$\pm$0.34[5]&--1.75$\pm$0.17[5]\\

\hline 
\end{tabular}
\tablefoot{This table is available entirety in a machine-readable form in the 
online journal.\\
Numbers in square brackets indicate the lines used to estimate the chemical abundances. Note that for stars with abundances based on single line, the dispersion gives the uncertainty on \feih\ measurement.}
\end{table*}
\begin{table*}
\footnotesize
\caption{Mean abundances, dispersions and confidence levels (C.L.) for data sets plotted in Fig.~\ref{ab-logg}}.
\label{tab_cl}
\centering
\begin{tabular}{cccc|cccccc}
\hline\hline
   &
 {\feh\ old}&
 {\feh\ int}&
 C.L. &
 {\mgh\ old}&
 {\mgh\ int}&
 C.L. \\
 \hline
 UVES & --2.31$\pm$0.27[10] & --1.81$\pm$0.17[34] & 82\% & --2.00$\pm$0.37[10] & --1.39$\pm$0.26[32] & 82\%\\ 
 GHR  & --2.19$\pm$0.27[4] & --1.69$\pm$0.18[4] & 84\% & --1.80$\pm$0.20[4] & --1.34$\pm$0.07[4] & 80\% \\  
 GMR & --1.93$\pm$0.16[9] & --1.54$\pm$0.23[10] & 75\% & --1.85$\pm$0.11[7] & --1.25$\pm$0.12[10] & 91\% \\
 ALL & --2.15$\pm$0.28[23] & --1.75$\pm$0.21[48] & 75\% & --1.91$\pm$0.22[21] & --1.35$\pm$0.22[46] & 83\% \\
\hline 
\end{tabular}
\tablefoot{Numbers in square brackets indicate the stars used to estimate the mean abundances.} 
\end{table*}


\section{Comparison with the Galactic halo}
\label{sec:abund_vs_mw}

\begin{table*}
\footnotesize
\caption{Comparison of mean abundances and dispersions between Carina, Halo and globular cluster stars.}
\label{tab_mean_compare}
\centering
\begin{tabular}{cccccccc}
\hline\hline
{Elem.} & \feh\ range & {Car. Old} & {Car. Int.} & {Carina} & {MW Halo} & {MW GCs} & {LMC GCs} \\
 \hline
\nafe & --1.95/--1.48 &\dots & 0.18$\pm$0.27[10] & 0.18$\pm$0.27[10] & --0.11$\pm$0.29[72]\tablefootmark{a} & 0.30$\pm$0.29[103]\tablefootmark{b} & 0.36$\pm$0.32[21]\tablefootmark{d} \\
\ofe & --2.60/--1.48 & 0.84$\pm$0.27[4] & 0.59$\pm$0.19[20] & 0.63$\pm$0.23[24] & 0.55$\pm$0.33[57]\tablefootmark{a} & 0.27$\pm$0.15[31]\tablefootmark{b} & 0.09$\pm$0.15[18]\tablefootmark{d} \\
\mgfe & --2.80/--1.18 & 0.29$\pm$0.28[21] & 0.40$\pm$0.22[46] & 0.36$\pm$0.24[67] & 0.34$\pm$0.19[581]\tablefootmark{a} & 0.27$\pm$0.12[139]\tablefootmark{b} & 0.09$\pm$0.22[21]\tablefootmark{d} \\
\sife & --2.14/--1.48 &1.09$\pm$0.12[1] & 0.54$\pm$0.22[16] & 0.56$\pm$0.25[17] & 0.27$\pm$0.25[87]\tablefootmark{a} & 0.32$\pm$0.10[60]\tablefootmark{b} & 0.38$\pm$0.15[20]\tablefootmark{d} \\
\cafe & --2.80/--1.48 &0.18$\pm$0.33[14] & 0.27$\pm$0.12[38] & 0.25$\pm$0.17[52] & 0.20$\pm$0.13[540]\tablefootmark{a} & 0.23$\pm$0.07[74]\tablefootmark{b} & 0.21$\pm$0.10[21]\tablefootmark{d} \\
\tiiife & --2.80/--1.48 & 0.40$\pm$0.24[9] & 0.24$\pm$0.28[33] & 0.28$\pm$0.30[42] & 0.34$\pm$0.15[515]\tablefootmark{a}& 0.35$\pm$0.19[11]\tablefootmark{c} & 0.53$\pm$0.45[3]\tablefootmark{d} \\
$[$Mg/Ca$]$ & --2.80/--1.48 & 0.13$\pm$0.23[14] & 0.15$\pm$0.20[36] & 0.15$\pm$0.21[50] & 0.03$\pm$0.17[534]\tablefootmark{a} & 0.03$\pm$0.07[74]\tablefootmark{b} & --0.12$\pm$0.29[21]\tablefootmark{d} \\
$\left[\frac{\rm Mg+Ca}{2\rm Fe}\right]$ & --2.80/--1.48 & 0.28$\pm$0.29[14] & 0.35$\pm$0.14[36] & 0.13$\pm$0.16[50] & 0.32$\pm$0.14[533]\tablefootmark{a} & 0.25$\pm$0.08[71]\tablefootmark{b} & 0.15$\pm$0.10[21]\tablefootmark{d} \\
$\left[\frac{\rm Mg+Ca+Ti}{3\rm Fe}\right]$ & --2.80/--1.48 & 0.34$\pm$0.30[9] & 0.32$\pm$0.17[32] & 0.33$\pm$0.19[41] & 0.33$\pm$0.13[506]\tablefootmark{a} & 0.29$\pm$0.12[15]\tablefootmark{c} & 0.18$\pm$0.09[21]\tablefootmark{d} \\
\hline
\end{tabular}
\tablefoot{Numbers in square brackets indicate the stars/GCs used to estimate the mean abundances.\\
\tablefoottext{a}{Individual MW halo dwarf/giant stars from \citet{frebel10mw}.}\\
\tablefoottext{b}{Individual stars for 19 Galactic globular clusters from \citet{carretta09uves,carretta09gir,carretta10uves}.}\\
\tablefoottext{c}{Mean abundances of Galactic globular clusters from \citet{pritzl05}.}\\\tablefoottext{d}{Individual stars for LMC globulars from \citet{mucciarelli10} and \citet{colucci12}.}\\

}

\end{table*}

\begin{figure}
\centering
\includegraphics[trim=0.5cm 0cm 4cm 0.5cm,clip,width=0.5\textwidth]{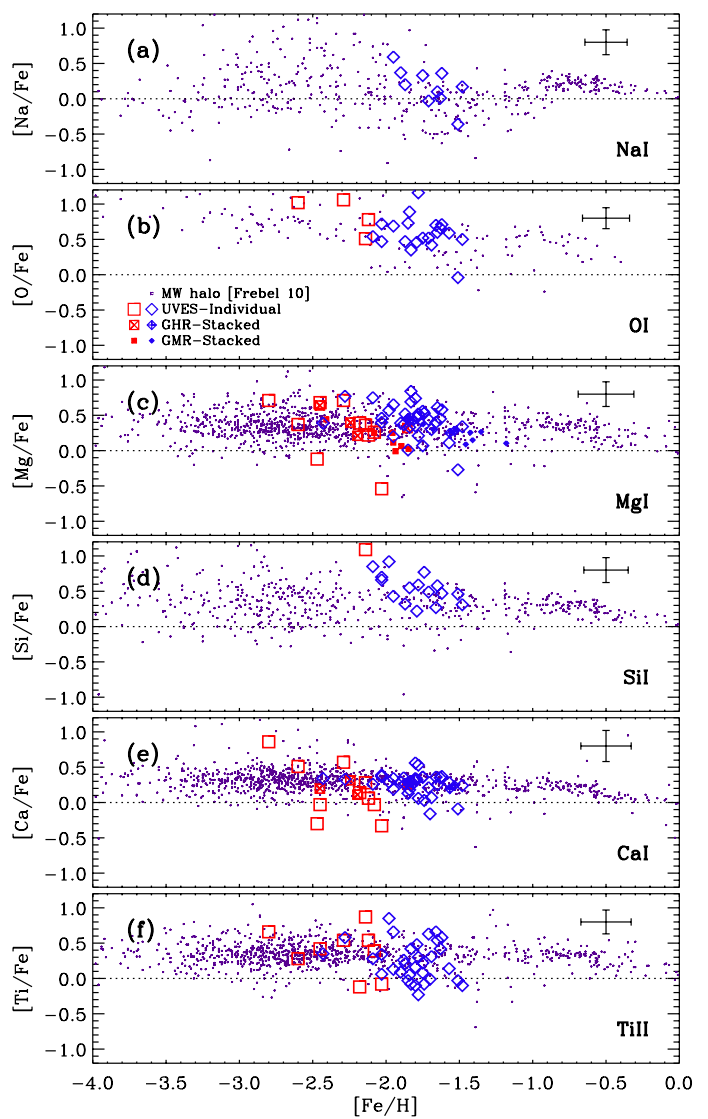}
\caption{Element abundances as function of \feh. The open red squares and blue diamonds
are the measurements based on UVES spectra of this work for the old and intermediate-age populations. 
The crossed squares and diamonds show the measurements based on Giraffe-HR spectra,
while small solid symbols are for the Giraffe-MR sample. 
The purple dots show the Milky Way halo stars from \citet{frebel10mw}. 
\label{mwelem}}
\end{figure}

Figure~\ref{mwelem} displays the abundance trends of five $\alpha$-elements,
including Na, 
for the entire sample of
old- (red squares) and intermediate-age (blue diamonds)
stars. For a detailed comparison with field halo stars, the
large sample of 
elemental abundances compiled by
\citet{frebel10mw} is shown as purple dots. These abundances are
based on high-resolution spectra of field stars of all
evolutionary stages. We note that these measurements have been
rescaled to the same solar elemental abundances adopted in this
investigation.\\ 

The [Na/Fe] abundances are only available for a limited sample
(ten) of intermediate-age stars. The mean weighted
abundance---[Na/Fe]=0.18 ($\sigma$=0.27)---appears slightly
larger than the abundances of
field halo stars in the iron range covered by 
Carina stars---[Na/Fe]=--0.11 ($\sigma$=0.29).
However, the difference is within 1$\sigma$ (see
Table~\ref{tab_mean_compare}). We note that the field value is
based on a large sample (72) and shows an intrinsic dispersion
that is higher than the individual measurements (see the error
bars plotted in the top right corner).
Moreover, intermediate-age Carina stars attain either solar or
slightly supersolar Na abundances. The \nafe\ abundances provided by
\citet{venn12} are on average subsolar. The discrepancy for the stars
with \feh$>$--2.0 is caused by the difference in the mean iron
abundance $\Delta$(our--Venn)=--0.37$\pm$0.11~dex (see Sect.~5.2 and 
Fig.~3 in Paper~V). In passing we note that the plausibility
of the current \nah\ abundances is supported by the mild difference with
similar abundances provided by \citet{shetrone03}, \citet{venn12}, and
\citet{koch08} (see panel (a) of Fig.~\ref{abcompare}).\\

[O/Fe] abundances are available for a few old (four) and for a
good sample of intermediate-age (20) stars. They are O enhanced
and attain very similar abundances within the errors (see
Table~\ref{tab_mean_compare}). The mean weighted [O/Fe] abundance
of the entire sample---[O/Fe]=0.63 ($\sigma$=0.23)---agrees quite
well with similar abundances---[O/Fe]=0.55 ($\sigma$=0.33)---for
field halo stars (57) in the same iron interval.
We note that for several metal-poor objects in our sample both O and Si
display very weak lines and their EWs have modest or poor precision.\\

[Mg/Fe] abundances are available for a sizable sample of both old
and intermediate-age stars (see Sect. \ref{sec:abund_vs_logg}). They
are Mg enhanced and agree---[Mg/Fe]=0.29 ($\sigma$=0.28) vs
[Mg/Fe]=0.40 ($\sigma$=0.22)---within the errors. We note that old
and intermediate-age Carina stars show more similar [Mg/Fe]
abundances than \mgh\ because the old sample is systematically
more iron-poor than the younger one. The mean
weighted [Mg/Fe] abundance of the entire sample---[Mg/Fe]=0.36
($\sigma$=0.24)---agrees very well with similar
abundances---[Mg/Fe]=0.34 ($\sigma$=0.19)---for field halo stars
(581) in the same metallicity interval. 
This finding supports early results obtained by \citet{idiart00} 
concerning the Mg abundances of field Halo stars. The 
non-LTE correction for the \mgi\ abundances of both halo and 
Carina stars were not taken into account. However, \citet{merle11} 
found that the non-LTE corrections to the EWs of two Mg lines at 
5711 and 5528~\AA\ are smaller than 10\%.\\ 

The [Si/Fe] abundances are available for a sizable sample of
intermediate-age (16) stars  but for only one old star. They are Si
enhanced and the mean weighted abundance of the entire
sample---[Si/Fe]=0.56 ($\sigma$=0.25)---is larger than the mean
abundance---[Si/Fe]=0.27 ($\sigma$=0.25)---of field halo stars
(87). They agree within 1$\sigma$. The mean Si abundance
decreases to 0.54~dex ($\sigma$=0.22) when the old star is excluded.\\ 

The [Ca/Fe] abundances of old (14) and intermediate-age (38)
Carina stars agree quite well---[Ca/Fe]=0.18 ($\sigma$=0.33) vs
[Ca/Fe]=0.27 ($\sigma$=0.12)---with each other. The weighted
mean [Ca/Fe] abundance of the entire sample---[Ca/Fe]=0.25
($\sigma$=0.17)---agrees very well with similar
abundances---[Ca/Fe]=0.20 ($\sigma$=0.13)---for field halo stars
(540) in the same iron interval. We excluded the non-LTE corrections 
to the EWs of \cai\ lines for both halo and Carina stars from
the comparison. \citet{merle11} found that the non-LTE corrections to the EWs of 
the two adopted \cai\ lines (6122, 6166~\AA) are smaller than 10\%.
The anonymous referee noted the paucity of subsolar [Mg/Fe] and
[Ca/Fe] abundance ratios, plotted in panels (c) and (e) of Fig.~\ref{mwelem}, 
when compared with similar abundances provided by \citet{lemasle12}. 
The good agreement between the two different data sets has already 
been discussed in Sect.~\ref{sec:abund}. The above difference is 
mainly caused by a difference of --0.27$\pm$0.09~dex in iron abundance.
We refer
to Paper~V for a more detailed discussion.\\

The [Ti/Fe] abundances are based on \tiii. The abundances of old
and intermediate-age Carina stars are enhanced and agree quite
well---[Ti/Fe]=0.40 ($\sigma$=0.24) vs [Ti/Fe]=0.24
($\sigma$=0.28). The former sample includes nine stars,
while the latter contains almost three dozen stars. The mean
weighted [Ti/Fe] abundance of the entire sample---[Ti/Fe]=0.28
($\sigma$=0.30)---agrees very well with similar
abundances---[Ti/Fe]=0.34 
($\sigma$=0.15)---for field halo stars
(515) in the same iron interval. 
The abundances for neutral \tii\ are not used here to
avoid non-LTE effects that cause an ionization imbalance in
this species, as shown by \citet{bergemann11} and
\citet{bergemann14}. It is noteworthy that the correction of
$+0.25$~dex for \tii, suggested by \citet{bergemann11} and based
on the metal-poor RGB star HD~122563 (\feh=--2.5), agrees very
well with the difference we found in our stars
\tii--\tiii=+0.28~dex.\\

\begin{figure}
\centering
\includegraphics[trim=0.5cm 17.5cm 4cm 0.5cm,clip,width=0.5\textwidth]{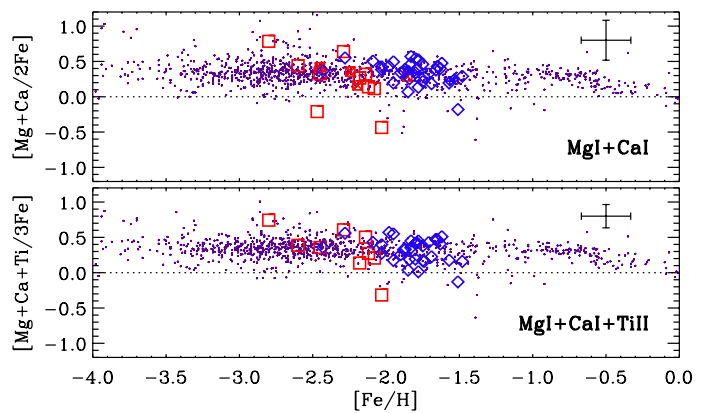}
\caption{Same as Fig.~\ref{mwelem}, but for the element
combination indicated. \label{mwalpha}}
\end{figure}

To further constrain the [$\alpha$/Fe] abundance of Carina stars,
we also summed the individual $\alpha$-elements with
reliable measurements. The top panel of Fig.~\ref{mwalpha} shows
[Mg+Ca/2Fe] as a function of the iron abundance. The
old and the intermediate-age subpopulations have, once again,
very similar abundances. They also agree quite well with similar
abundances for field halo stars (see also
Table~\ref{tab_mean_compare}). The same result is found for the
[Mg+Ca+Ti/3Fe] $\alpha$-element abundances plotted in
the bottom panel of that figure. 
The standard deviations of the Carina subpopulations are, as noted
by the anonymous referee, larger than the standard deviations of the
halo sample. The difference is mainly due to the sample size. We
performed a number of tests and found that
the Mg distribution of Carina and halo stars agree at 95\%\ CL. 
We found a similar agreement for the Ca (90\%\ CL) distribution, 
while for Ti it is at 50\%\ CL.
These findings are soundly supported by the mean of the 
$\alpha$-elements plotted in Fig.~\ref{mwalpha} and listed in
Table~\ref{tab_mean_compare}. The sum of Mg and Ca do agree at 
99\%\ CL, while the sum of the three $\alpha$-elements (bottom panel of 
Fig.~\ref{mwalpha}) agree at 75\%\ CL.

This comparison highlights two relevant findings.\\
{\it (i)} The [$\alpha$/Fe] abundances of old and intermediate-age
Carina stars are enhanced. They do not show
any significant difference within the errors.\\ 
{\it (ii)} The current mean weighted [$\alpha$/Fe] abundances
agree quite well with similar abundances of field halo stars in
the same range in iron as covered by Carina RG stars.

\section{Comparison with globular clusters}
\label{sec:abund_vs_GC}

The comparison between Carina's 
elemental abundances and 
abundances in the Galactic halo is partially hampered by the fact that 
the latter abundances are derived from 
spectra with different 
spectral resolutions and different wavelength ranges. To further constrain 
the $\alpha$-element abundances of Carina stars, we repeated the
comparison using abundances of RG stars in Galactic
\citep{pritzl05,carretta09uves,carretta09gir,carretta10uves} and
Magellanic \citep{mucciarelli10,colucci12} globular clusters. 

This sample has several distinct differences compared to the field stars: 
{\it (i)} a significant fraction of the abundances rely on high-resolution spectra
similar to those of the Carina stars. They also cover very similar 
wavelength ranges and therefore similar line lists.
{\it (ii)} A significant fraction of the abundances are on a homogenous 
$\alpha$-element scale.
{\it (iii)} The spectroscopic targets include only cluster RG stars.
{\it (iv)} They show distinctive spectroscopic features (anticorrelations) when 
compared with field stars, thus suggesting a different chemical enrichment history.
 
\begin{figure}
\centering
\includegraphics[trim=.5cm 0cm 4cm 0.5cm,clip,width=0.5\textwidth]{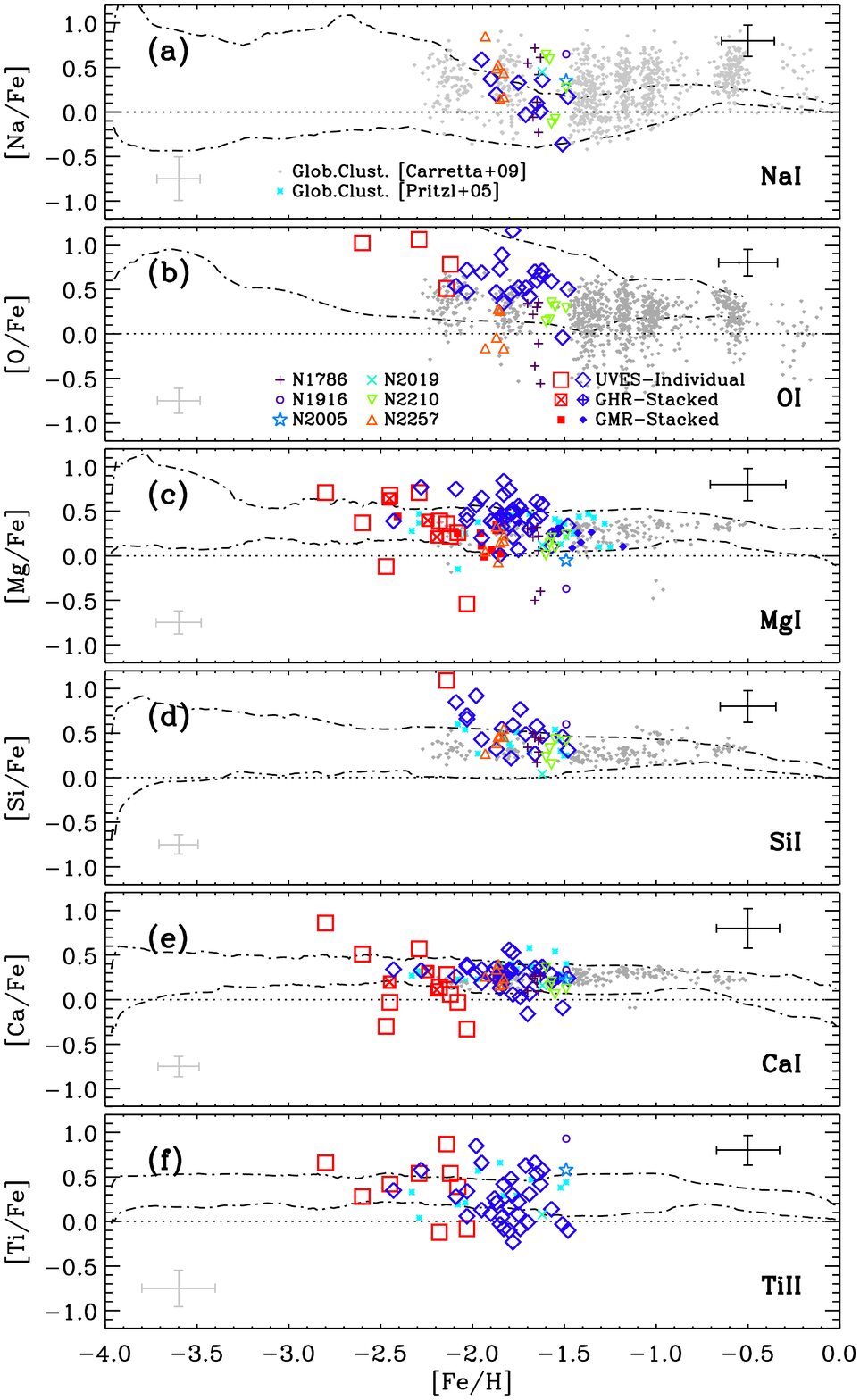}
\caption{Element abundances as functions of \feih\ for Galactic and some LMC
globular clusters. Red squares and blue diamonds show abundances of
old and intermediate-age Carina stars. The small cyan asterisks are mean
abundances for Galactic globular clusters from \citet{pritzl05}. 
The colored small symbols are data for LMC clusters from 
\citet[][NGC~1786, NGC~2210, and NGC~2257]{mucciarelli10} and 
\citet[][NGC~1916, NGC~2005, and NGC~2019]{colucci12}.
The gray dots are individual
abundances for 19 Galactic globular clusters from
\citet{carretta09uves,carretta09gir,carretta10uves}. 
The gray error bars in the bottom left corner of each
panel show the mean abundance errors in GCs. The two dot-dashed lines
show the limiting positions of the Milky Way halo stars \citet{frebel10mw}.
\label{gcelem}} 
\end{figure}

Panel (a) of Fig.~\ref{gcelem} shows that the Na abundances of
Carina's intermediate-age RGs agree quite well with cluster stars.
However, Carina RGs, in the metallicity range they cover, 
attain Na abundances that are slightly underabundant compared to the
cluster abundances. They appear, indeed, to agree better
with the Na abundances of field halo stars (see Table~\ref{tab_mean_compare}).
The two dot-dashed lines plotted in Fig.~\ref{gcelem} display the limiting
position of Milky Way halo stars according to \citet{frebel10mw}. To avoid
spurious fluctuations in the range of elemental abundances covered by
field stars, we ranked the entire sample as a function of the iron
abundance. Then we estimated the running average by using a box including
the first 100 objects in the list. We estimated the mean abundances
(iron, element) and the standard deviations of the subsample. We
estimated the same quantities by moving one object in the ranked list
until we took account of the last 100 objects in the sample. We
performed several tests changing both the number of objects included in
the box and the number of stepping stars. We found that the limiting
positions are minimally affected by plausible variations.\\
The comparison between Carina and cluster O abundances is shown in panel (b) 
of the same figure. Here, the situation is reversed: 
they attain O abundances that are slightly enhanced 
compared with cluster stars. The (anti-)correlation Na--O of Carina stars 
is discussed in more detail in Sect. \ref{sec:na-o}.\\
The Mg abundances of Carina RGs agree quite well with cluster Mg abundances. 
They show, within the errors, very similar enhancements over the 
entire metallicity range covered by both globular and Carina samples. \\
The same conclusion applies to globular and Carina Si abundances 
(see panel (d) of Fig.~\ref{gcelem}).\\
The comparison between globular and Carina Ca abundances appears to be more complex. 
Panel (e) of Fig.~\ref{gcelem} shows that Carina's intermediate-age 
subpopulation agrees quite well with globular Ca abundances. 
On the other hand, Carina's old subpopulation shows a slightly broader
spread when compared with cluster stars (see Table~\ref{tab_mean_compare})
and with the
intermediate-age subpopulation. The internal difference appears
reliable ($\sigma$=0.33 vs 0.12~dex), since it is differential and based
on GHR and UVES spectra. However, more accurate Ca abundances of Carina
old-population stars are required to confirm this preliminary
evidence.
In passing we note that the current findings 
support previous results by \citet{thevenin01} for Mg and Ca abundances 
of seven turn-off stars in the metal-poor Galactic globular cluster NGC~6397.\\
The bottom panel of Fig.~\ref{gcelem} shows the comparison between globular 
cluster and Carina \tiii\ abundances. The two samples agree quite well over 
the entire metallicity range. There is mild evidence that a fraction of 
Carina stars might be slightly underabundant in \tiii\ for \feh=--1.8, 
but the difference is within the intrinsic dispersion of the two samples 
(see error bars).  

\begin{figure}
\centering
\includegraphics[trim=0.5cm 17.5cm 4cm 0.5cm,clip,width=0.5\textwidth]{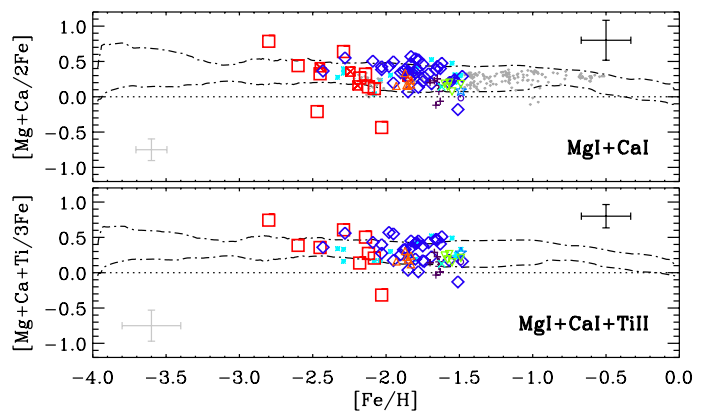}
\caption{Same as Fig.~\ref{gcelem}, but for the element
combination indicated. \label{gcalpha}}
\end{figure}

The top and bottom panels of Fig.~\ref{gcalpha} reveal to
even a cursory scrutiny that the sum of Mg and Ca and
the sum of Mg, Ca, and \tiii\ agree quite well with the mean $\alpha$-element 
abundances of globular stars. This indicates that the 
$\alpha$-element enrichments appear to be quite similar. This 
evidence is quite compelling because it applies not only to the old, 
but also to the intermediate-age subpopulation. 
In passing we note that this comparison also suggests 
that nearby stellar systems and field halo stars attain very similar $\alpha$ enhancements
in the metallicity range they cover. 
This further supports the evidence that $\alpha$-elements, in contrast 
with {\it s}- and {\it r}-elements, are poor diagnostics to 
constrain possible differences in chemical enrichment between old 
and intermediate-age stellar populations \citep{cescutti08,matteucci14}.

\section{Comparison with nearby dwarfs}
\label{sec:abund_vs_dwarf}

\begin{table*}
\footnotesize
\caption{Mean abundances and dispersions for dSph and UFD galaxies.}
\label{tab_dwarf}
\centering
\begin{tabular}{ccccccc}
\hline\hline
{Elem.} & Draco\tablefootmark{a} 
            & Fornax\tablefootmark{b} 
            & LeoI\tablefootmark{c} 
            & Sculptor\tablefootmark{d}  
            & Sextans\tablefootmark{e} 
            & UrsaMinor\tablefootmark{f}  \\
 \hline
\feih & --2.13$\pm$0.57[15] &
--1.27$\pm$0.55[388]&
--1.22$\pm$0.20[2] &
--2.27$\pm$0.97[17] &
--2.53$\pm$0.52[14] &
--1.83$\pm$0.30[6] \\

\naife & --0.19$\pm$0.33[15] &
--0.41$\pm$0.28[84] &
--0.31$\pm$0.23[2] &
--0.15$\pm$0.35[17] &
--0.04$\pm$0.28[14] &
--0.01$\pm$0.41[6] \\

\oife & 0.38$\pm$0.29[11] &
0.57$\pm$0.76[3] &
0.35$\pm$0.24[2] &
0.92$\pm$1.38[11] &
0.33$\pm$0.43[7] &
0.28$\pm$0.18[6] \\

\mgife & 0.13$\pm$0.26[15] &
--0.05$\pm$0.15[201]&
--0.06$\pm$0.24[2] &
0.19$\pm$0.26[17] &
0.09$\pm$0.26[14] &
0.30$\pm$0.25[6] \\

\siife & --0.56$\pm$1.48[13] &
--0.03$\pm$0.18[223] &
--0.03$\pm$0.22[2] &
0.52$\pm$0.91[11] &
0.24$\pm$0.22[7] &
0.45$\pm$0.37[6] \\

\caife & 0.05$\pm$0.15[15] &
--0.22$\pm$0.13[84] &
0.06$\pm$0.21[2] &
0.12$\pm$0.27[17] &
0.13$\pm$0.16[14] &
0.12$\pm$0.16[6] \\

\tiiife & 0.39$\pm$0.31[11] &
0.14$\pm$0.21[220] &
0.25$\pm$0.24[2] &
0.18$\pm$0.31[13] &
0.09$\pm$0.27[9] &
\dots \\

\hline
{Elem.} & Bo\"otes\tablefootmark{g} 
             & UrsaMaj\tablefootmark{h} 
             & ComaBer\tablefootmark{i} 
             & Hercules\tablefootmark{j} 
             & LeoIV\tablefootmark{k} \\
\hline
\feih & --2.35$\pm$0.29[7] &
--2.89$\pm$0.52[3] &
--2.57$\pm$0.30[3] &
--2.03$\pm$0.34[2] &
--3.14$\pm$0.27[1] \\

\naife & \dots&
--0.06$\pm$0.09[3] &
--0.24$\pm$0.49[3] &
0.70$\pm$0.22[2] &
0.03$\pm$0.36[1] \\

\oife & \dots&
1.60$\pm$0.94[3] &
1.00$\pm$0.55[3] &
1.01$\pm$0.24[2] &
 \dots\\

\mgife & 0.46$\pm$0.14[7] &
0.56$\pm$0.19[3] &
0.55$\pm$0.43[3] &
0.79$\pm$0.20[2] &
0.34$\pm$0.25[1] \\

\siife &  \dots&
1.24$\pm$0.32[3] &
0.71$\pm$0.36[3] &
0.60$\pm$0.20[2] &
 \dots \\

\caife & 0.26$\pm$0.14[7] &
0.37$\pm$0.17[3] &
0.45$\pm$0.32[3] &
0.03$\pm$0.21[2] &
0.23$\pm$0.22[1] \\

\tiiife & \dots&
0.19$\pm$0.05[3] &
0.23$\pm$0.23[3] &
 \dots&
0.38$\pm$0.35[1] \\

\hline
\end{tabular}
\tablefoot{Numbers in square brackets indicates the stars used to estimate the mean abundances.\\
\tablefoottext{a}{\citet{shetrone01,fulbright04,cohen09}} --
\tablefoottext{b}{\citet{shetrone03,tafelmeyer10,letarte10,hendricks14}} --
\tablefoottext{c}{\citet{shetrone03}} --
\tablefoottext{d}{\citet{shetrone03,geisler05,frebel10scl,starkenburg13}} --
\tablefoottext{e}{\citet{tafelmeyer10,aoki09,shetrone01}} --
\tablefoottext{f}{\citet{shetrone01}} --
\tablefoottext{g}{\citet{feltzing09,norris10}} --
\tablefoottext{h}{\citet{frebel10ufd}} --
\tablefoottext{i}{\citet{frebel10ufd}} --
\tablefoottext{j}{\citet{koch08her}} --
\tablefoottext{k}{\citet{simon10}}
}
\end{table*}

\begin{figure}
\centering
\includegraphics[trim=0.5cm 0cm 4cm 0.5cm,clip,width=0.5\textwidth]{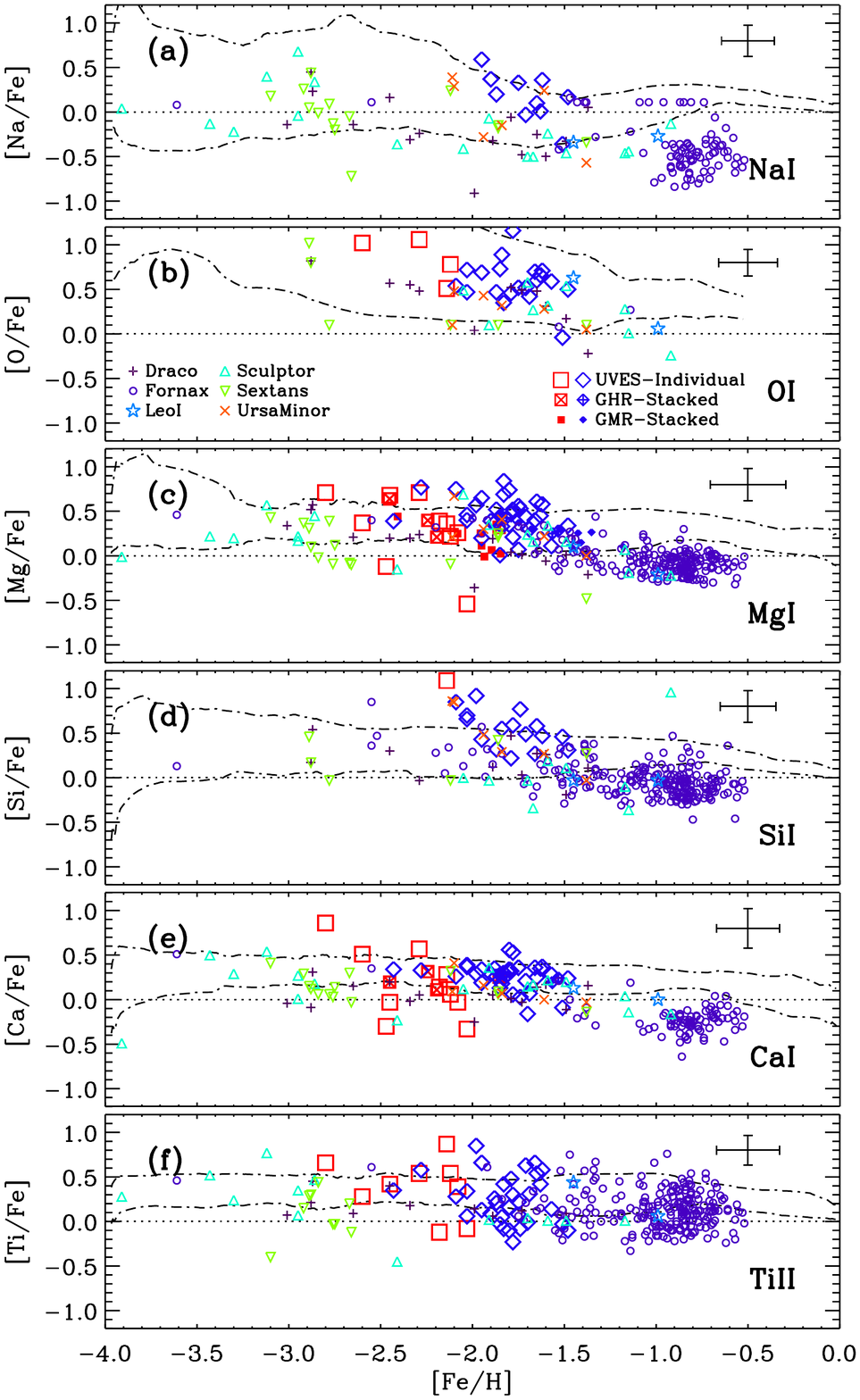}
\caption{Element abundances as functions of \feih\ for the dwarf spheroidal galaxies
\citep[Draco:][pluses]{shetrone01,fulbright04,cohen09} -
\citep[Fornax:][circles]{shetrone03,tafelmeyer10,letarte10,hendricks14} -
\citep[LeoI     :][stars]{shetrone03} -
\citep[Sculptor:][triangles]{shetrone03,geisler05,frebel10scl,starkenburg13} -
\citep[Sextans:][upside-down triangles]{shetrone01,aoki09,tafelmeyer10} -
\citep[Ursa Minor:][crosses]{shetrone01}.
\label{dwarfelem}}
\end{figure}

\begin{figure}
\centering
\includegraphics[trim=0.5cm 0cm 4cm 0.5cm,clip,width=0.5\textwidth]{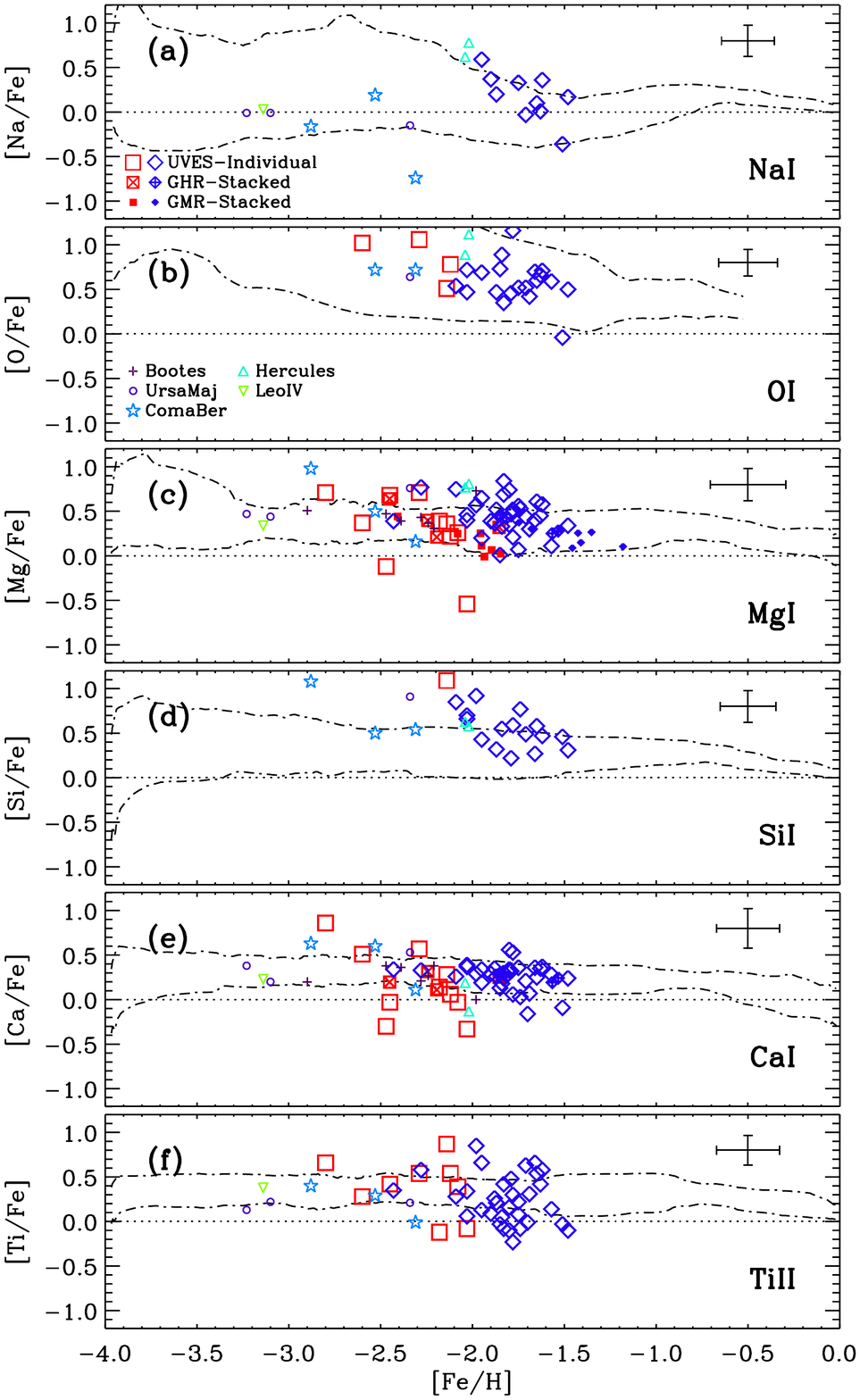}
\caption{$\alpha$-element abundances as functions of \feih\ for the ultra-faint dwarf galaxies:
\citep[Bootes:][pluses]{feltzing09,norris10} -
\citep[Ursa Major:][circles]{frebel10ufd} -
\citep[ComaBer:][stars]{frebel10ufd} -
\citep[Hercules:][triangles]{koch08her} -
\citep[LeoIV:][upside-down triangles]{simon10}.
\label{ufdelem}}
\end{figure}

To further characterize the chemical enrichment history of Carina's old 
and intermediate-age subpopulations, we extended the comparison 
to other nearby dSphs and UFDs. 
The dSphs included in the current comparison---Draco,
Fornax, LeoI, Sculptor, Sextans, and Ursa Minor---have accurate 
elemental abundances from high-resolution spectra, 
covering a broad range in iron abundances (see Table~\ref{tab_dwarf}).
Moreover, they show quite different star formation histories, 
but they all host a clearly defined old (t$\sim$12~Gyr) subpopulation. 
Panels (a) and (b) of Fig.~\ref{dwarfelem}
display the comparison between 
Na and O abundance in Carina and the 
selected dSphs. 
These data show that Na and O abundances in 
nearby dSphs agree within the errors with abundances in 
field halo stars over the entire metallicity range covered by 
dSphs. The only exception is Fornax. This is the most metal-rich 
system and has Na abundances (purple asterisks) that are systematically 
lower by $\sim$0.3-0.5~dex than field halo stars and 
the few metal-rich stars in Sculptor (cyan triangles). A similar 
underabundance in Na was also found by \citet{mcwilliam13} in 
RGs of the metal-rich Sagittarius dSph galaxy. 
This is a metallicity regime in which Na abundances might be affected 
by non-LTE effects \citep{gratton99,carretta10m54}, but the 
detailed spectroscopic analysis performed by \citet{fulbright07}
among K-type 
giants and FGK-type dwarfs in the Galactic disk indicates that 
the non-LTE effects are weak (see also \citealt{mcwilliam13}). 

Panels (c), (d), and (e) show the comparison between Mg, Si, and Ca 
abundances in Carina and other nearby dwarfs. Stars in dSphs are all 
enhanced in these elements and agree with each other over 
the entire metallicity range. They also agree quite well with 
abundances in field Halo stars (dashed lines). 
The only exception is, once again, Fornax, showing a well-defined 
underabundance in the quoted $\alpha$ elements. There are a few 
metal-rich stars in Sculptor showing mild underabundances, but the 
possible difference is within 1$\sigma$.  
The bottom panel (f) shows that \tiii\ abundances
in nearby dSphs are on average enhanced over the entire metallicity 
range. Moreover, they agree quite well with each other and with 
field Halo stars. The same agreement is also found for
Fornax stars. There is weak evidence that the dispersion in \tiii\
abundances is, at fixed metal content, slightly higher in dwarfs 
than in the field (see also dispersion values listed in 
Table~\ref{tab_mean_compare} and \ref{tab_dwarf}).  

The insight emerging from this comparisons does not 
allow us to reach firm conclusions concerning the chemical enrichment 
history of Carina and nearby dwarfs. Indeed, O, Mg and Na are mainly produced by 
massive stars during hydrostatic burning phases, and they appear to have 
similar abundances in nearby dSphs and among field halo stars. 
On the other hand, the most metal-rich systems (Fornax and Sagittarius) 
appear to be underabundant in these three elements. 
The scenario becomes even more surprising for the explosive $\alpha$-elements, 
namely Si, Ca, and Ti. Si and Ca abundances in field halo stars and 
in nearby dwarfs, except for Fornax, agree quite well. Once again, metal-rich systems 
show either solar or slightly underabundant Si and Ca abundances.    
On the other hand, Ti abundances agree quite well over the entire metallicity 
range covered by the nearby dSphs. 

We performed the same comparisons with RGs in five nearby UFDs
(Bo\"otes, Ursa Major, Coma Ber, Hercules, and Leo IV) in
Fig.~\ref{ufdelem}. The results are similar to the results
found for metal-por dSphs (see Fig.~\ref{dwarfelem} and Table~\ref{tab_dwarf}). 
However, the sample of stars is still too limited to reach firm conclusions. 

In conclusion, we are left with the following empirical evidence:
$\alpha$-element abundances in nearby dwarf are similar to the Galactic
field halo stars and to globular clusters in the metal-poor regime
(\feh$<$--1.5). The difference is smaller on average than 1$\sigma$ (see
Tables~\ref{tab_mean_compare} and \ref{tab_dwarf}). 
There is change in the trend when moving into the more metal-rich regime
(\feh$>$--1.5). The Fornax \afe\ abundance ratios are on average
underabundant when compared with halo stars. Sculptor appears to be
a transitional stellar system, since the \afe\ abundance ratios are
slightly higher or lower than solar.

\begin{figure}
\centering
\includegraphics[trim=0.5cm 17.5cm 4cm 0.5cm,clip,width=0.5\textwidth]{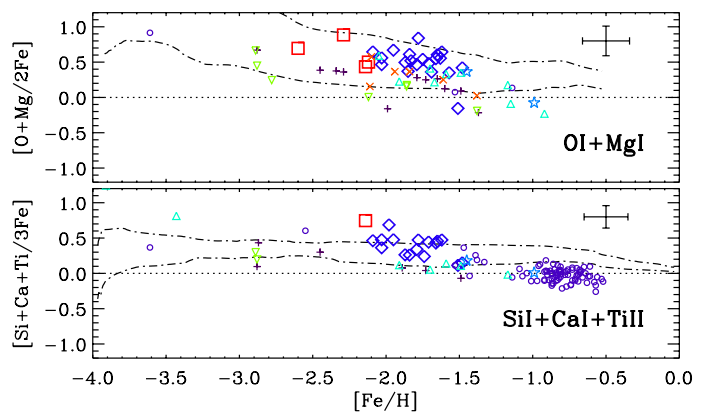}
\caption{Same as Fig.~\ref{dwarfelem}, but for the element 
combination indicated. \label{dwarfalpha}}
\end{figure}

\subsection{Hydrostatic vs. explosive}

To further investigate the difference between hydrostatic and explosive
elements, the top panel of Fig.~\ref{dwarfalpha} shows the comparison
between the sum of Mg and O for field halo stars. In particular, Mg is
produced in hydrostatic core C and O burning, while Ca is one product of
explosive Si burning during the supernova type II (SN~II) explosion.
They overlap quite well until the metal-rich regime. The bottom panel
shows the comparison of the sum of the explosive $\alpha$-elements (Si,
Ca, and Ti). The agreement is quite good in the metal-poor and in the
metal-intermediate iron regimes. 
The depletion of the quoted sum for Fornax stars in the metal-rich
regime is somehow mitigated by the inclusion of titanium. The
depletion might have been even stronger if we had only summed Si and Ca
abundances of Fornax stars.

\begin{figure}
\centering
\includegraphics[trim=0.5cm 13.7cm 4.cm .5cm,clip,width=0.5\textwidth]{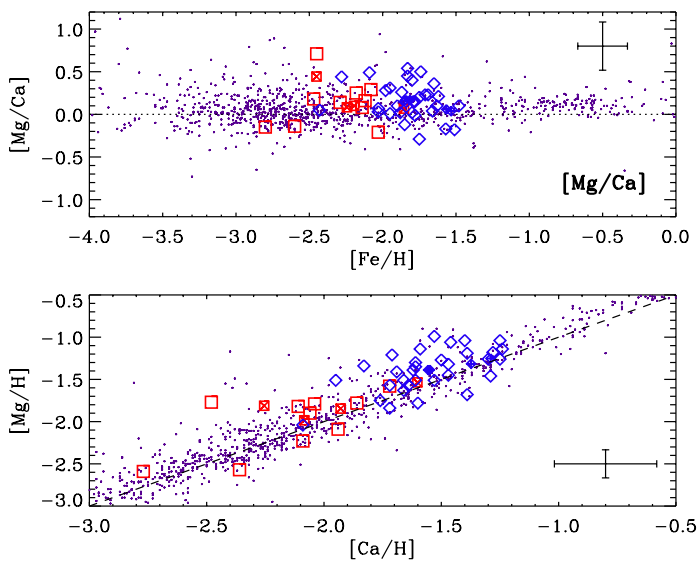}
\caption{Top panel: the ratio [Mg/Ca] as a function of \feh
for Carina as compared to MW halo stars. Symbols and colors are
the same as in Fig.~\ref{mwelem}. The abundances of \mgh\ and
\cah\ are compared in the bottom panel, where the dashed line
shows the bisector of the plane. 
\label{mgca}} 
\end{figure}

Figure~\ref{mgca} shows the comparison between the abundances of two
elements, Mg and Ca, as yields of SN~II events. The
different ratios of these elements are due to the progenitor mass of the
SN~II \citep{iwamoto99}. For Carina, the ratio [Mg/Ca] shows
a weaker enhancement than in the MW stars (0.15 vs 0.03~dex, top
panel, see also Table~\ref{tab_mean_compare}), but  it is well within 1$\sigma$
(0.27~dex). The same behavior is found in the comparison between individual
abundances of \mgh\ and \cah\ (bottom panel).

\section{Carina chemical enrichment}
\label{sec:na-o}

Data plotted in Figs.~\ref{mwelem} and \ref{mwalpha}
show that Carina's chemical
enrichment history is quite complex. Similar conclusions were also
reached by \citet{lemasle12} and \citet{venn12}, who found
evidence that the metal-poor subpopulation is less $\alpha$-enhanced
than the metal-rich one. This result was independently
supported by \citet{deboer14}, who performed a detailed star formation
history of the Carina dSph galaxy.     
On the other hand, the current individual (Fig.~\ref{mwelem}) 
and mean (Fig.~\ref{mwalpha}) 
$\alpha$ abundance ratios of the two subpopulations are very
similar within 
1$\sigma$. Data listed in Table~\ref{tab_mean_compare} 
indicate that the difference is at most on the order of 0.1~dex. 
However, \mgfe\ (panel (c) Fig.~\ref{mwelem}) and \cafe\ (panel (e) Fig.~\ref{mwelem})
abundance ratios of the old subpopulation appear to be less
$\alpha$-enhanced than the intermediate-age subpopulation in the iron
range (--2.3$<$\feh$<$--1.9) they have in common (see also the top panel of
Fig.~\ref{mwalpha}). The comparison for the other $\alpha$-elements is hampered by
statistics and by the limited range in iron abundance in common between
the two subpopulations.   

The mean $\alpha$-abundance ratios plotted
in the bottom panel of Fig.~\ref{mwalpha} show that the sum of Mg, Ca, and Ti does
not display any significant difference between the old- and the
intermediate-age subpopulation. The main difference between the current
analysis and previous investigations available in the literature is in
the sample size. We worked with $\alpha$-element abundances for 67
stars, 46 of which belonged to the intermediate-age subpopulation.
The sample discussed by \citet{lemasle12} is a factor of two smaller
(35 objects). The difference in the sample size becomes on the order of
20\%\ (55 objects) if we also include abundances on high-resolution
spectra provided by \citet{shetrone03}, \citet{koch08}, and
\citet{venn12}.  

This evidence indicates that homogeneous $\alpha$-element
abundances for a sizable sample of RGB stars do not show a clear
difference between old- and intermediate-age subpopulations. The same
outcome applies to the possible occurrence of a "knee" either in the
metal-poor (\feh=--2.5) or in the metal-rich subpopulation (\feh=--1.6).
There are three (Car45, Car27,  and Car19) stars in the top panel and two
(Car27 and Car19) in the bottom panel of Fig.~\ref{mwalpha} that
show less 
enhanced $\alpha$ abundance ratios. However, the difference is either 
within or slightly larger than 1$\sigma$.

To further constrain the chemical enrichment history of Carina,
we also investigated the (anti-) correlation between Na and
O. There is solid evidence that evolved and unevolved cluster
stars display a well-defined anticorrelation in Na--O and in Mg--Al 
\citep{carretta09uves,carretta09gir,carretta14}.
We note that the environment appears to play a minor role, if any, in these 
cluster star anticorrelations, and indeed, they have also been identified 
in globulars belonging to LG dwarf galaxies (LMC, \citealt{mucciarelli10}; 
Fornax, \citealt{letarte06}).

The occurrence of light-element anticorrelations in GCs is considered
to be the consequence of deep potential wells that are able to retain the
ejecta of candidate stellar polluters, such as intermediate-mass
asymptotic giant branch stars and/or fast-rotating massive stars (see
\citealt{cassisi13book} and references therein). Nearby dwarf galaxies
typically have low central stellar densities
\citep{mateo98araa,mcconnachie12}, therefore a correlation between Na
and O is expected. However, we still lack detailed spectroscopic
investigations of nearby dSphs that are characterized by high central densities
(LeoI, Draco, Ursa Minor). Accurate light element abundances in these
systems are required before reaching firm conclusions concerning the
environmental impact on their chemical enrichment history.

\begin{figure}
\centering
\includegraphics[trim=0.5cm 13cm 11.5cm 0.5cm,clip,width=0.4\textwidth]{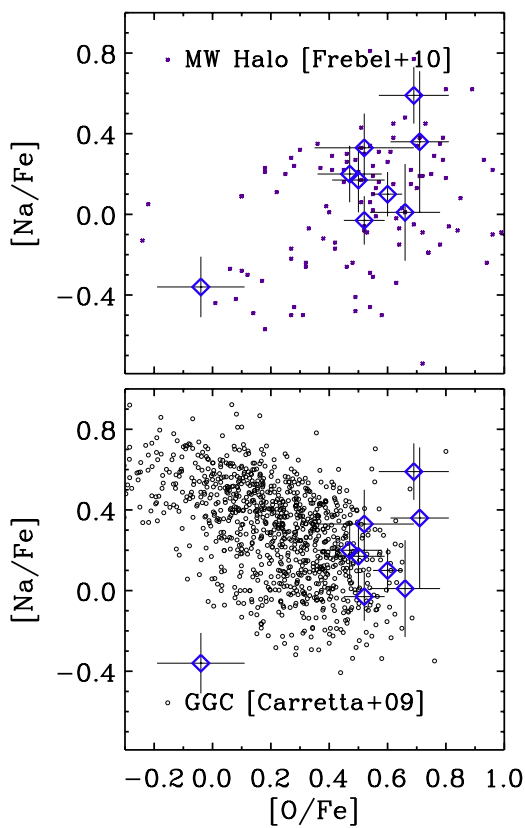}
\caption{Comparison of Carina stars with MW halo stars (top)
and with 1958 stars of 19 galactic globular clusters
\citep[bottom,][]{carretta09uves,carretta09gir} in the classical Na vs O diagram.
\label{nao}} 
\end{figure}

The top panel of Fig.~\ref{nao} shows that Carina stars have a
(positive) correlation between Na and O. Moreover, the correlation is
quite similar to the correlation of field halo stars found by
\citet{frebel10mw}. The current data soundly support previous results
obtained by \citet{carretta10m54} and \citet{mcwilliam13} for
Sagittarius stars. The key advantage of the current comparison is that
we investigate the correlation for a system that is significantly
more metal-poor than Sagittarius ($\sim-2.0$ vs $\sim-0.6$~dex). To
define the difference with cluster stars on a more quantitative basis,
the bottom panel shows the comparison between the current sample and the
entire sample of cluster stars investigated by
\citet{carretta09uves,carretta09gir}. The difference is quite clear, and
indeed Carina stars display a steady increase in the regime of \ofe\
abundances in which the \nafe\ in Galactic globulars becomes less and
less abundant. Unfortunately, we cannot constrain whether the candidate
old stars show the same trend, since the Na abundance measurements for
those stars are lacking.

\section{Summary and final remarks}
\label{sec:summary}

We have presented a new spectroscopic investigation of Carina RG stars. The
abundance analysis was focused on Na plus five $\alpha$-elements:
O, Mg, Si, Ca, and Ti. The current approach, when compared with similar
spectroscopic investigations available in the literature, has two
distinct features.\\
{\it (i)} We used spectroscopic data collected with UVES (high spectral
resolution) and with \giraffe\ (high- and medium-resolution) at the VLT.
The current spectroscopic data sets cover a significant fraction of
Carina's RGB and, for the first time, reach the red clump stars
(\vv$\sim$20.5, \bmv=0.6 mag), that is, a reliable tracer of the intermediate-age
stellar population. We obtained accurate abundance analyses
for 44 RGs based on UVES spectra that were previously analyzed in the literature
\citep{koch08,venn12,fabrizio12}. They were supplemented with 65 
(high-resolution, \citealt{lemasle12,fabrizio11}) and 802 
(medium-resolution, 
\citealt{koch06,fabrizio11}) \giraffe\ spectra. The
abundance analysis of 46\%\ of the former sample and 84\%\ of the
latter were discussed here for the first time.\\
{\it (ii)} We took advantage of the new photometry index \cubi\
introduced by \citet{monelli13,monelli14} as an age and probably 
a metallicity indicator to split stars along the Carina's RGB.
It is noteworthy that the main conclusion of this investigation, that is, the presence of 
two subpopulations that experienced two different chemical enrichment 
histories, is not affected by the intrinsic parameters affecting the \cubi\
index.

To improve the accuracy of the abundance analysis in the faint
magnitude limit, we devised a new data reduction strategy. The
entire \giraffe\ data set was divided into ten surface gravity and
effective temperature bins. The spectra of the stars belonging to
the same gravity and temperature bin are characterized by similar stellar
parameters and were stacked together. 
This allowed us to increase the 
signal-to-noise ratio in the faint magnitude limit (\vv$\ge$20.5~mag)
by at least a factor of five.
In this context we note that the spectra of the stars 
belonging to the same gravity and temperature bin are quite similar 
because of the modest variation in the intrinsic parameters. This means 
an improvement in the accuracy of individual abundance estimates. Moreover, 
this approach allowed us to control possible systematics (surface gravity 
and effective temperature dependence of non-LTE effects) between the old 
and intermediate-age stellar populations.

On the basis of these data sets, we have performed the
largest and the most homogeneous abundance analysis of the Carina dSph
galaxy. The abundances were estimated using both EWs (high-resolution
spectra) and spectrum synthesis (medium-resolution stacked spectra).

The main results of the current analysis are listed below.\\
$\bullet$ There is increasing evidence that Carina's old and
intermediate-age stellar populations display two distinct
\feh\ and \mgh\ distributions. The dichotomy
is present over the entire gravity range (0.5$<$\logg$<$2.5); this
means from the tip of the RGB down to the RC stars. Specifically,
we found that the old stellar populations has a mean iron abundance
of --2.15$\pm$0.06~dex ($\sigma$=0.27), while the intermediate-age population 
has a mean iron abundance of --1.75$\pm$0.03~dex ($\sigma$=0.21). The two 
distributions differ at the 75\%\ level. This agrees quite well with preliminary
results by \citet{monelli14} based on data available in the literature
and with \citet{lemasle12}, using a subsample of the current spectroscopic data set.
Moreover, we found that the old and intermediate-age
stellar populations have mean \mgh\ abundances of 
--1.91$\pm$0.05~dex ($\sigma$=0.22)
and of --1.35$\pm$0.03~dex ($\sigma$=0.22). They differ at the 83\%\ level.\\
$\bullet$ The individual \afe\ abundances of Carina's old and intermediate-age
evolved stars are enhanced over the entire iron range.\\
$\bullet$ Carina's $\alpha$-element abundances and abundances
for Galactic halo stars agree quite well (1$\sigma$) over the entire 
iron range covered by Carina stars. The same conclusion applies to the 
comparison between $\alpha$-element abundances in Carina and in Galactic 
and Magellanic globular clusters. However, Na and O abundances display 
different trends.\\
$\bullet$ Carina's $\alpha$-element abundances also agree within
1$\sigma$ with similar abundances for LG dwarf spheroidals and 
ultra-faint dwarf galaxies in the iron range we considered.\\
$\bullet$ We found evidence of a clear correlation between
Na and O abundances. Carina's correlation agrees quite well with
the typical Na--O correlation of MW halo stars. This supports
previous findings by \citet{carretta10m54} and 
\citet{mcwilliam05}. 

These results support the evidence of a close similarity in 
the chemical enrichment history of field halo and Carina stars
\citep{idiart00}. 

The stacked spectra will also allow us to investigate the abundances
of several {\it s}- and {\it r}-elements. 
Of course, the data reduction we devised
to stack the spectra in gravity and temperature bins is opening the path
to a detailed spectroscopic investigation of the old HB stars
(\vv$\sim$21-21.5~mag), the most reliable tracers of the Carina old stellar
population.

\begin{acknowledgements}

It is a pleasure to acknowledge the anonymous referee
for her/his comments and suggestions that improved the content and 
the readability of our manuscript.
M.F. acknowledges financial support from the PO FSE Abruzzo 2007-2013
through the grant "Spectro-photometric characterization of stellar populations
in Local Group dwarf galaxies" prot.89/2014/OACTe/D (PI:~S.~Cassisi). 
He also thanks ESO for support as science visitor (DGDF12-42, F.~Primas).
This work was partially supported by PRIN-INAF 2011 "Tracing the formation and 
evolution of the Galactic halo with VST" (P.I.: M. Marconi), by PRIN-MIUR 
(2010LY5N2T) "Chemical and dynamical evolution of the Milky Way and Local Group 
galaxies" (P.I.: F. Matteucci) and by the Education and Science Ministry of
Spain (grant AYA2010-16717). S.C. and R.B. warmly thank for the financial support from 
PRIN-INAF 2014 "The kaleidoscope of stellar populations in Galactic Globular Clusters 
with Hubble Space Telescope" (PI: S. Cassisi).

\end{acknowledgements}

\bibliography{biblio}


\end{document}